\documentclass[structabstract]{aa}
\pdfoutput=1
\usepackage{txfonts,graphicx}
\usepackage{natbib,twoopt}
\usepackage[breaklinks=true]{hyperref} 
\usepackage[usenames,dvipsnames,svgnames,table]{xcolor}

\bibpunct{(}{)}{;}{a}{}{,} 
\makeatletter
\newcommandtwoopt{\citeads}[3][][]{\href{http://adsabs.harvard.edu/abs/#3}%
{\def\hyper@linkstart##1##2{}%
\let\hyper@linkend\@empty\citealp[#1][#2]{#3}}}
\newcommandtwoopt{\citepads}[3][][]{\href{http://adsabs.harvard.edu/abs/#3}%
{\def\hyper@linkstart##1##2{}%
\let\hyper@linkend\@empty\citep[#1][#2]{#3}}}
\newcommandtwoopt{\citetads}[3][][]{\href{http://adsabs.harvard.edu/abs/#3}%
{\def\hyper@linkstart##1##2{}%
\let\hyper@linkend\@empty\citet[#1][#2]{#3}}}
\newcommandtwoopt{\citeyearads}[3][][]%
{\href{http://adsabs.harvard.edu/abs/#3}
{\def\hyper@linkstart##1##2{}%
\let\hyper@linkend\@empty\citeyear[#1][#2]{#3}}}
\newcommandtwoopt{\citedoi}[3][][]{\href{http://dx.doi.org/#3}%
{\def\hyper@linkstart##1##2{}%
\let\hyper@linkend\@empty\citealp[#1][#2]{#3}}}
\newcommandtwoopt{\citepdoi}[3][][]{\href{http://dx.doi.org/#3}%
{\def\hyper@linkstart##1##2{}%
\let\hyper@linkend\@empty\citep[#1][#2]{#3}}}
\newcommandtwoopt{\citetdoi}[3][][]{\href{http://dx.doi.org/#3}%
{\def\hyper@linkstart##1##2{}%
\let\hyper@linkend\@empty\citet[#1][#2]{#3}}}
\newcommandtwoopt{\citeyeardoi}[3][][]%
{\href{http://dx.doi.org/#3}
{\def\hyper@linkstart##1##2{}%
\let\hyper@linkend\@empty\citeyear[#1][#2]{#3}}}
\newcommandtwoopt{\citegallica}[3][][]{\href{http://gallica.bnf.fr/ark:/12148/#3}%
{\def\hyper@linkstart##1##2{}%
\let\hyper@linkend\@empty\citealp[#1][#2]{#3}}}
\newcommandtwoopt{\citepgallica}[3][][]{\href{http://gallica.bnf.fr/ark:/12148/#3}%
{\def\hyper@linkstart##1##2{}%
\let\hyper@linkend\@empty\citep[#1][#2]{#3}}}
\newcommandtwoopt{\citetgallica}[3][][]{\href{http://gallica.bnf.fr/ark:/12148/#3}%
{\def\hyper@linkstart##1##2{}%
\let\hyper@linkend\@empty\citet[#1][#2]{#3}}}
\newcommandtwoopt{\citewmo}[3][][]{\href{https://www.wmo.int/pages/prog/arep/gaw/ghg/documents/#3}%
{\def\hyper@linkstart##1##2{}%
\let\hyper@linkend\@empty\citealp[#1][#2]{#3}}}
\newcommandtwoopt{\citepwmo}[3][][]{\href{https://www.wmo.int/pages/prog/arep/gaw/ghg/documents/#3}%
{\def\hyper@linkstart##1##2{}%
\let\hyper@linkend\@empty\citep[#1][#2]{#3}}}
\newcommandtwoopt{\citetwmo}[3][][]{\href{https://www.wmo.int/pages/prog/arep/gaw/ghg/documents/#3}%
{\def\hyper@linkstart##1##2{}%
\let\hyper@linkend\@empty\citet[#1][#2]{#3}}}
\makeatother

\begin{document}

\title{\texttt{Molecfit}: A general tool
  for telluric absorption correction\thanks{\texttt{Molecfit} is available at
      \url{http://www.eso.org/pipelines/skytools}}}
\subtitle{I. Method and application to ESO instruments\thanks{Based  on observations  made with  ESO  Telescopes at  the La  Silla
  Paranal Observatory under programme IDs:  
60.A-9100, 
60.A-9452, 
076.C-0129, 
075.A-0603, 
079.D-0374, 
080.D-0526, 
290.C-5022, 
092.D-0024, 
084.D-0912, 
085.D-0161, 
086.D-0066, 
and
088.D-0109 
}
}

\author{
A.~Smette\inst{1}
\and
H.~Sana\inst{1,2}
\and
S. Noll\inst{3}
\and
H.~Horst\inst{1,4}
\and
W.~Kausch\inst{3,5}
\and
S.~Kimeswenger\inst{6,3}
\and
M.~Barden\inst{7}
\and 
C.~Szyszka\inst{3}
\and
A.~M.~Jones\inst{3}
\and
A.~Gallenne\inst{8}
\and
J.~Vinther\inst{9}
\and
P.~Ballester\inst{9}
\and 
J.~Taylor\inst{9}
}

\institute{
European Southern Observatory,
Casilla 19001,
Alonso de Cordova 3107
Vitacura, Santiago, 
Chile\\
\email{asmette@eso.org}
\and
now at ESA / Space Telescope Science Institute,
3700 San Martin Dr, Baltimore, 
MD 21218, United States
\and
Institute for Astro and Particle Physics, Universit{\"{a}}t Innsbruck,
Technikerstrasse 25, 6020 Innsbruck, Austria
\and
Josef-Führer-Straße 33, 80997 M\"unchen, Germany
\and
University of Vienna, Department of Astrophysics, Türkenschanzstr. 17
(Sternwarte), A-1180 Vienna, Austria
\and
Instituto de Astronom{\'{i}}a, Universidad Cat{\'{o}}lica del Norte,
Avenida Angamos 0610, Antofagasta, Chile
\and
International Graduate School of Science and Engineering, Technische Universit\"at M\"unchen,
Boltzmannstr. 17, D-85748 Garching bei M\"unchen, Germany
\and
Universidad de Concepción,
Casilla 160-C, Concepción, Chile
\and
European Southern Observatory,
Karl-Schwarzschild-Strasse 2,
85748 Garching bei München,
Germany
}

\date{Received <date> /
Accepted <date>}

\abstract{ 
  The  interaction of  the light  from astronomical  objects  with the
  constituents  of  the   Earth's  atmosphere  leads  to  the
  formation  of  telluric  absorption  lines  in  ground-based
    collected  spectra.  Correcting  for these  lines, mostly
    affecting the  red and infrared region of  the spectrum, usually
  relies on observations of  specific stars obtained close in
  time and  airmass to the  science targets, therefore  using precious
  observing time.  
}
{ 
  We  present  \texttt{molecfit},  a  tool  for  correcting  for  telluric
  absorption  lines based  on synthetic  modelling of  the
  Earth's atmospheric
  transmission. \texttt{Molecfit}  is versatile  and can be  used with
  data  obtained  with  various ground-based  telescopes  and
  instruments.
}
{ 
  \texttt{Molecfit}  combines a publicly  available radiative
  transfer  code,  a   molecular  line  database,  atmospheric
    profiles,  and various  kernels to  model  the instrument
  line spread function. The atmospheric profiles are created by merging
  a standard atmospheric  profile representative of a
    given  observatory's  climate,   of local  meteorological  data,  and
  of dynamically  retrieved altitude   profiles for
  temperature, pressure, and  humidity. We discuss the
    various   ingredients  of  the   method,  its   applicability,  and
its    limitations. We also show examples of telluric line correction on
  spectra  obtained with a  suite of  ESO \emph{Very
    Large Telescope (VLT)} instruments.
}
{ Compared to  previous similar tools, \texttt{molecfit} takes  the best  results for  temperature, pressure,  and
    humidity in  the atmosphere above  the observatory
    into  account. As  a result,
  the standard deviation of the residuals after correction of
  unsaturated telluric  lines is frequently better than  2\% of the
  continuum.}
{  \texttt{Molecfit}  is able  to  accurately  model  and correct  for
  telluric  lines  over a  broad  range  of  wavelengths and  spectral
  resolutions. The  accuracy reached is comparable to or  better than the
  typical   accuracy   achieved  using   a   telluric  standard   star
  observation.  The availability  of such  a general  tool for
    telluric  absorption correction  may improve  future observational
    and  analysing strategies,  as well  as empower  users  of archival
    data.  }

\maketitle

\section{Introduction}
\label{sec:introduction}

Ground-based  observations  and, in  particular,  spectroscopy can  be
strongly  affected by absorption  caused by  molecules in  the Earth's
atmosphere.   The wavelength ranges  involved include, but  are not
limited to, the near- and mid-infrared. 
It is a standard practice to
correct  as much as possible for such telluric absorption lines in order to recover  
the spectrum  (or photometry) of  the target as it would
appear if the instrument was located above the atmosphere.  
Different approaches have been proposed.

A very  common method is to observe  either a rapidly
  rotating  early-type  or a G-type
star  (depending on  the specific  purpose) soon  before or  after the
observations of the science target. This \textit{\emph{telluric}} star should
ideally be  located close to the  science target in  the sky.  Indeed,
early type stars are advantageous because they are mostly featureless,
except for hydrogen lines.  Spectral regions close in wavelength to the
one of hydrogen  lines, on the other hand, are therefore best corrected
by  G-type stars.   The  reason  to observe  both  science target  and
telluric star  close in time is  to eliminate
differences caused by temporal variation in the atmosphere as much  as the possible.  On the other hand,
the need  to observe  both objects angularly  close on the  sky arises
because  saturated and unsaturared telluric absorption  lines  do not
scale with the same function of airmass. In addition, the distribution
of molecules in the atmosphere  (mainly  water vapour) can depend on
the pointing direction.

Observations of such telluric standard stars require telescope
  time  that   could  otherwise  be  dedicated   to  other  scientific
  observations.  This  is a necessary but  costly practice, especially
  in the coming era of extremely large telescopes.  Whenever a limited
  time  window   of  observability  is   available,  or  in   the case  of
  high-cadence  time series,  observations of  telluric  stars further
  reduce,  sometimes considerably,  the  amount of  time  that can  be
  dedicated to the main science targets.

The quality of the correction can  be variable and depends considerably on the
difference  in   airmass  between   the  science  and   telluric  star
observations,  because in  the best  conditions,  a difference  of 1\%  in
airmass  already introduces  a  1\%  change in  the  optical depth  of
unsaturated telluric  lines.  Examples  of  other problems
  affecting   observations  of  telluric   stars  are that
(i)  suitable  telluric stars that are  close  enough  to the  science
  target may not be  available;  (ii) atmospheric conditions
may   change   by  the   end   of   the  science   exposures;
(iii)  the  signal-to-noise  ratio  of  the  obtained
  telluric spectrum may be insufficient, either in comparison
  with  the quality  of the  science observations  or with  respect to
  one's  scientific aims;  (iv) post-observation
  analysis of the spectrum  of the telluric star reveals that
  it    does    not   have    the    expected    spectral   type    or
  characteristics. Critically, telluric star observations may
  be missing  as a result of inclement  weather, instrumental issues,
  human  errors,  and/or  incomplete  calibration plan.   The  use  of
archival  data provides  a particularly  illustrative example
  since  observations  of  a  telluric   star  may  be  missing  or  not
appropriate for the science goal pursued by the archive user as it may
significantly differ from the one originally intended.

Alternatively,  the emergence of  freely available  radiative transfer
codes  for  atmospheric  research  provides  the  possibility  of  using
synthetic transmission  spectra instead of telluric  standard stars.  One
of the first  articles to demonstrate the ability of  this approach is
\citetads{2007PASP..119..228B}, who used  the radiative transfer model
SMART     \citepads{1996JGR...101.4595M}     for     that     purpose.
\citetads{2010A&A...524A..11S}  refined the  method  incorporating the
radiative transfer code LBLRTM \citep{CLO05}, the line database HITRAN
\citep{ROT09},  a  combination  of  meteorological data  from  various
sources to  achieve synthetic transmission  curves, and a model  of the
line spread function.  These were then used to successfully
perform     telluric      absorption     corrections     of     CRIRES
\citepads{2004SPIE.5492.1218K}  spectra up to  an accuracy  of
about 2 per cent.     The     code    LBLRTM     is    also     used    by
\citetads{2013arXiv1312.2450H}  who incorporated it into their own
code to fit an entire input    spectrum.     

A     different    approach    is    used    by
\citetads{2014MNRAS.439..387C}.   Since  they  investigate  solar  system
planetary  atmospheres  containing similar  molecules to those in the  Earth's
atmosphere, they  derive the state  of the latter by  fitting spectral
features of telluric standard star observations. \cite{GAR13} used the
radiative  transfer model\footnote{RTM  is  part of  the SCIATRAN  2.2
  software package, developed by Institute of Remote Sensing/Institute
  of  Environmental  Physics of  the  University  of Bremen,  Germany}
developed for  the SCIAMACHY instrument onboard  the ENVISAT satellite
to fit  water features visible in  spectra taken at  the Sierra Nevada
Observatory.  In  addition, they  used it to  determine the  amount of
precipitable water  vapour in  the Earth's atmosphere  at the  time of
observation.

Unfortunately, the  limited quality of  these
corrections,   the   specialised   design  and/or   restricted
  availability has  so far impeded  any wide-spread use of  such tools.
The deviation  of the modelled  absorption line from the  observed one
can  be caused  by errors  in the  radiative transfer  modelling, poor
representation  of  the line  spread  function, and/or  limited
  accuracy of  the meteorological data. In  practice, therefore, a
general tool allowing the use of simulated atmosphere transmission
  spectra to correct  for Earth's atmosphere telluric lines  in a wide
  variety  of  data  is  simply  not available  to  the  astronomical
community at large.

Here  we describe\texttt{ molecfit}, a  tool for  modelling telluric
lines.   \texttt{ Molecfit} retrieves  the  most appropriate  atmospheric
profile (i.e., the variation in the temperature, pressure, and humidity
as  a  function  of  altitude)  for  the time  of  the  given  science
observations. It uses a state-of-the-art radiative transfer modelling
of the  Earth's atmosphere, together  with a comprehensive  database of
molecular parameters.   Finally, it allows for  fitting
  an analytical line spread function, as well as using a
  user-provided numerical  kernel.  \texttt{ Molecfit} can be  used from
0.3 to 30  $\mu$m (or more) and can  be configured to
  apply on  data collected  by a wide-range  of optical  and infrared
  instruments from most observatories, making it very versatile.

The initial  development of \texttt{ molecfit} was  carried out mostly
as a  set of IDL  routines used as  drivers for the  Reference Forward
Model\footnote{\url{http://www.atm.ox.ac.uk/RFM/}}  (hereafter, RFM)
radiative transfer
code.   The  initial  goal  was  to  measure  the  amount  of
precipitable  water vapour towards  zenith (hereafter,  PWV)
using the $\approx$  19.5 $\mu$m emission line with  VISIR, the mid-infrared
spectrograph  and imager  at  the VLT  \citepads{2008eic..work..433S}.
The code was then slightly modified to measure the amount of PWV using
sky emission line spectra  obtained with the cryogenic high-resolution
infrared  echelle spectrograph CRIRES,  since the  5.038 to  5.063 $\mu$m
range  is purely  dominated by  water vapour\footnote{These  VISIR and
  CRIRES measurements, together with UVES and X-shooter measurements
  obtained   using   a   different   method,   are   available   at
  \url{http://www.eso.org/observing/dfo/quality/GENERAL/PWV/HEALTH/trend_report_ambient_PWV_closeup_HC.html}.}.

The good quality  of the fits indicated that  generalising the code to
absorption lines and other spectral ranges was promising and led to a
well-developed  IDL   prototype  \citepads{2010HiA....15..533S}.   Its
performances    were     similar    to    the     one    created    by
\citetads{2010A&A...524A..11S} that  only focused on  CRIRES spectra.
This prototype  was then  ported in a  robust way  to CPL\footnote{The
  Common                        Pipeline                       Library
  \url{http://www.eso.org/sci/software/cpl/documentation.html}    based
  on  ANSI C.}  and  further developed  and optimised  as part  of the
Austrian in-kind  contribution for joining  ESO.  In the  process, the
LNFL/LBLRTM   code   \citep{CLO05}   was   chosen  instead   of   RFM,
because it (i) was more frequently updated at the time of the code
  selection,   (ii) is  used   by   a  large   community,  (iii) incorporates   a
  code-optimised  and  up-to-date  line  list,  and (iv)  comprises  broader
  functionality (e.g.,  UV/optical ozone absorption,  which was needed
  in a  parallel project \citepads{2012A&A...543A..92N})  and performs significantly
  faster for  most spectral ranges  for similar or  better precision.
This series of papers describes the results of these efforts.
In  the  present work,  we  describe  the  key ingredients  of
  \texttt{molecfit}, its applicability, and  limitations, and we show a
variety of applications  for ESO/VLT data covering a  range of spectral
resolving  power  and  wavelength   regions.   In  a  separate  paper,
\citep[][Paper  II]{KAUSCH14},  we have systematically  investigated
  the quality  of the telluric line absorption  correction by applying
  it to the large sample of archival VLT/X-shooter near-infrared data,
  and compared it with the standard method performed with the IRAF {\tt
    telluric}\footnote{\url{http://iraf.net/irafhelp.php?val=telluric&help=Help+Page}}
  task.

This  paper  is  organised  as follows.   Section~\ref{sec:atmoabsorb}
describes the  most important  aspects regarding the  absorption lines
formed  in  the  Earth's  atmosphere.   Section~\ref{sec:description}
provides a description of the \texttt{molecfit} package.
A  few  characteristic cases  are  described  in
Sect.~\ref{sec:usage}.   Limitations  to the  use  of  the tool  are
summarised  in Sect.~\ref{sec:limitations}. Examples  of successful
applications of  \texttt{molecfit} on  spectra obtained with  several ESO
instruments  are  given   in  Sect.~\ref{sec:examples}.  The most
important aspects of \texttt{molecfit} are summarised in the conclusion.
The technical aspects of \texttt{molecfit} 
and additional user guidance, including a description of the 
Graphical User Interface (GUI) that allows for 
data interaction, are provided in the User Manual \citep{NOL13},
(hereafter, UM).

\section{Absorption arising in the Earth's atmosphere}
\label{sec:atmoabsorb}

\begin{figure*}
\centering
\includegraphics[clip=true]{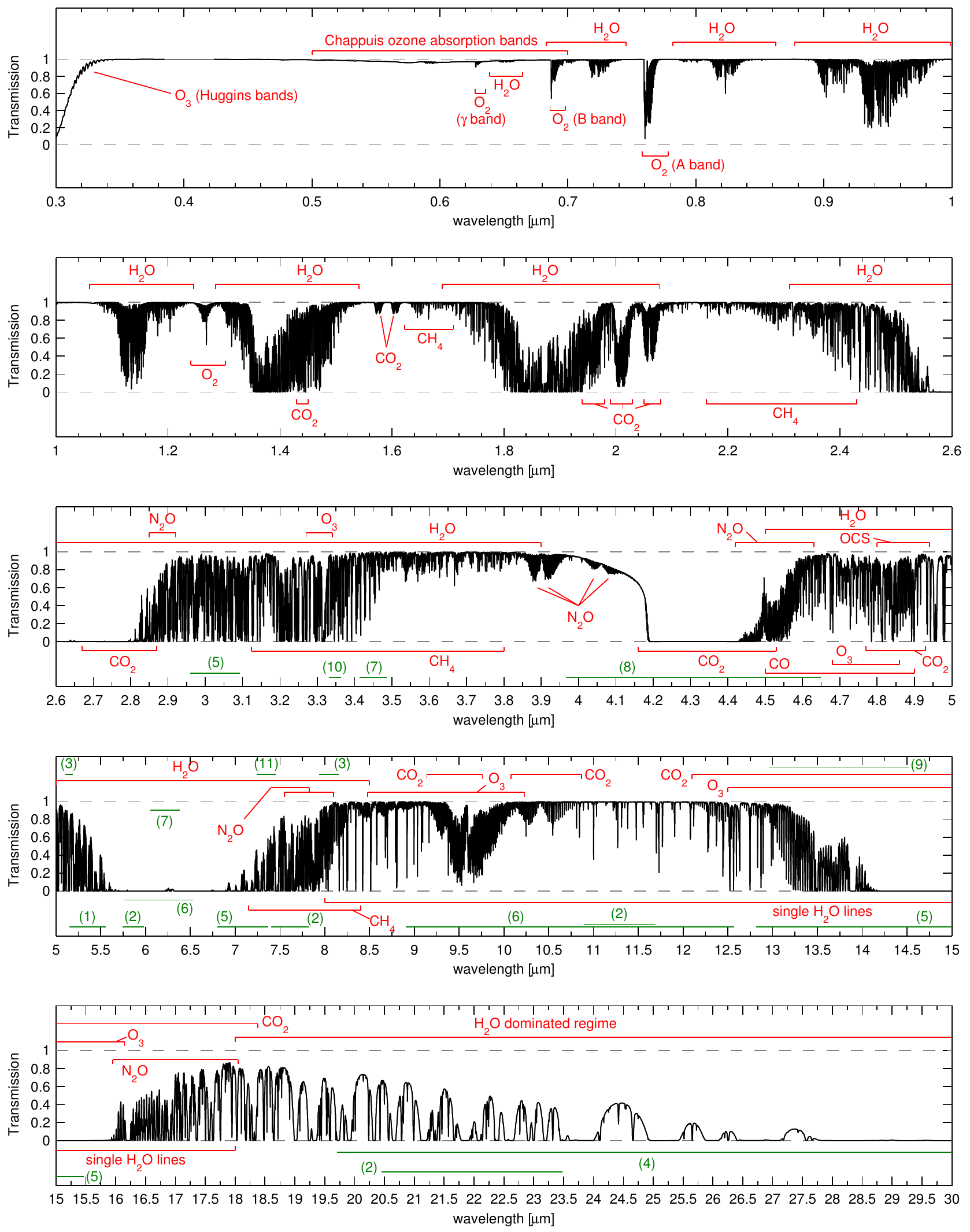}
\caption[]{Synthetic absorption  spectrum of  the sky between  0.3 and
  30~$\mu$m calculated with LBLRTM (resolution $R \sim 10\,000$) using the
  annual  mean profile for  Cerro Paranal  \citep{2012A&A...543A..92N}.  The eight main
  molecules O$_2$, O$_3$, H$_2$O,  CO, CO$_2$, CH$_4$, OCS, and N$_2$O
  contribute  more  than 5\%  to  the  absorption  in some  wavelength
  regimes.  The red  regions mark the ranges where  they mainly affect
  the  transmission, minor  contributions of  these molecules  are not
  shown.    The  green   regions  denote   minor   contributions  (see
  Table~\ref{tab:tracegases})  from the  following molecules:  (1) NO,
  (2) HNO$_3$, (3)  COF$_2$, (4) H$_2$O$_2$, (5) HCN,  (6) NH$_3$, (7)
  NO$_2$,  (8)  N$_2$,  (9)  C$_2$H$_2$, (10),  C$_2$H$_6$,  and  (11)
  SO$_2$.}
\label{fig:trans}
\end{figure*}

The Earth's  atmosphere consists of  about $78\%$ of N$_2$,  $21\%$ of
O$_2$,  $1\%$ of Ar,  and several trace  gases and  aerosols. Each  of the
molecules  and  aerosols  affects  the  light  travelling  through  in
different      wavelength      regimes      by     absorption      and
scattering.  Figure~\ref{fig:trans} shows  a model  transmission curve
derived  with   the  sky  model   developed  in  a   parallel  project
\citepads{2012A&A...543A..92N}.   The main  absorption  regions of  the eight  major
contributing  molecular  species  O$_2$,  O$_3$, H$_2$O,  CO,  CO$_2$,
CH$_4$,  OCS, and  N$_2$O are  marked. Thanks  to the  complexity  of the
molecules, several ro-vibrational bands are clearly visible.

The optical regime  is dominated by broad absorption  bands from ozone
(Huggins  bands  shortwards  of  $\lambda  \sim  0.4  \,  \mu\mathrm{m}$,
\citetads{1890SidM....9..318H};   Chappuis  bands   at  $\sim   0.5  <
\lambda/\mu\mathrm{m}  <  0.7  $,  \citetgallica{bpt6k30485}),  narrow
oxygen  bands (the prominent  A, B  and the  weaker $\gamma$  bands at
$\sim   0.759   <   \lambda/\mu\mathrm{m}   <  0.772$,   $   0.686   <
\lambda/\mu\mathrm{m} <  0.695$, and $0.628  < \lambda/\mu\mathrm{m} <
0.634  $,  respectively),  and   some  comparably  weak  water  vapour
features.  The  latter become dominant  in the entire  infrared regime
longwards  of the  $I$  band.  In  addition  in the  $JHK$ regime,  the
transmission is  significantly affected by CO$_2$,  CH$_4$, and O$_2$.
Between the $K$ band and $\sim18\,\mu\mathrm{m}$ absorptions are mainly
caused  by H$_2$O  and  CO$_2$, with  some  contribution from  N$_2$O,
CH$_4$, and  minor absorption  features by CO  and OCS. In  the regime
from 18  to 30\,$\mu$m only water  vapour is dominant.  Over the whole
wavelength range,  the gases listed  in Table~\ref{tab:tracegases} lead
to absorption  features with optical depth  reaching up to  5\% at the
given spectra resolving power ($R \sim 10\,000$).

The Earth's  atmosphere is highly  variable on several time  scales in
temperature,  pressure, and chemical  composition.  The  most variable
relevant molecule  is water vapour,  which is directly related  to the
actual  weather  conditions at  the  time  of observations.   However,
seasonal, daily,  and sometimes hourly variations  in the volume mixing ratio
of  various  gases  have  been  measured, in  particular,  for  CO$_2$
\citepdoi{10.1029/JD094iD06p08549},                              CH$_4$
\citepdoi{10.5194/acp-9-443-2009},              and              O$_3$
\citepdoi{10.1038/172633a0}.    In  addition,   following   the  World
Meteorological Organization\footnote{\url{http://www.wmo.int/gaw}}, the
fraction  of CO$_2$  has increased  by $10-15 $,  CH$_4$  by $\sim10$,  and
N$_2$O       by      $\sim8$              since      1985
\citepwmo{GHG_Bulletin_No.8_en.pdf} leading  to global warming.  Daily
columns for a number of  molecules (O$_3$, N$_2$O, CH$_2$, etc.) 
over most of the Earth surface can be obtained from \url{http://www.temis.nl/}.

\begin{table}
\caption[]{Species with a minor line-absorption contribution. The
specific spectra are calculated with the radiative transfer code
LBLRTM (resolution $R \sim 10\,000$) without continua to extract only the line contribution. The wavelength ranges are also shown in Fig.~\ref{fig:trans}.}
\centering
\vspace{5pt}
\begin{tabular}{c c c c}
\hline\hline
\noalign{\smallskip}
Species & Number in & Wavelength range & Optical\\
 & Fig.~\ref{fig:trans} & [$\mu$m] & depth [\%]\\
\noalign{\smallskip}
\hline
\noalign{\smallskip}
NO          & 1  & 5.153--5.556    & $\la$0.3 \\
HNO$_3$     & 2  & 5.743--5.983    & $\la$1 \\
            &    & 7.404--7.825    & $\la$2 \\
            &    & 10.890--11.694  & $\la$1 \\
            &    & 20.452--23.478  & $\la$3 \\
COF$_2$     & 3  & 5.103--5.187    & $\la$0.01\\
            &    & 7.935--8.157    & $\la$0.01\\
H$_2$O$_2$  & 4  & > 19.705         & $\la$0.3\\
HCN         & 5  & 2.959--3.092    & $\la$1\\
            &    & 6.801--7.370    & $\la$1\\
            &    & 12.819--15.467  & $\la$4\\
NH$_3$      & 6  & 5.755--6.530    & $\la$0.01\\
            &    & 8.904--12.571   & $\la$1\\
NO$_2$      & 7  & 3.412--3.487    & $\la$0.1\\
            &    & 6.051--6.383    & $\la$4\\
N$_2$       & 8  & 3.966--4.650    & $\la$2\\
C$_2$H$_2$  & 9  & 12.957--14.523  & $\la$0.1\\
C$_2$H$_6$  & 10 & 3.332--3.364    & $\la$2\\
SO$_2$      & 11 & 7.243--7.455    & $\la$0.01\\
\noalign{\smallskip}
\hline
\end{tabular}
\label{tab:tracegases}
\end{table}

\section{Description of the \texttt{molecfit} telluric correction tool}
\label{sec:description}

The  \texttt{molecfit}   package  allows  one  to   simulate  and  fit
tropospheric and  stratospheric emission or  absorption telluric lines
affecting  user-selected region(s) of  an observed  (usually, science)
spectrum. It uses parameters provided  in a configuration file and, if
present,  the  FITS  header  of  the  spectrum.   It  then  returns  a
transmission spectrum for the whole wavelength range of the spectrum.

\texttt{Molecfit} includes the following ingredients:

\begin{enumerate}
\item an automatic  retrieval of two atmospheric  profiles  from the
 Global Data   Assimilation System (GDAS)  web site.
  These profiles bracket the  time of the observation and can
  be  retrieved  for  any  observatory  in  the  world. These  profiles 
  are linearly interpolated to  match the
  time  of the  observations and  merged with  a standard  profile and
  local  meteorological  data  to  create  an input  profile  for  the
  radiative transfer code;
\item a  radiative transfer code that simulates  atmospheric emission and
  transmission spectra; 
\item a  molecular spectroscopic  database;
\item a simple grey-body calculator that simulates the contribution
  of the continuum from the instrument and telescope;
\item a choice of possible line spread functions (also referred
  to as \emph{\emph{kernels}});
\item  an  automatised  fitting  algorithm  that  adjusts  the  continuum
  spectrum,  wavelength calibration,  spectral resolution,  and column
  density of each  of the relevant molecules, with  the possibility of
  excluding spectral regions expressed either in pixels or in wavelength
  regions.   

\end{enumerate}


In the following, we  describe important aspects of these components 
in more detail. We also discuss  specific choices made for the ESO 
Paranal observatory.

\subsection{Atmospheric profile}\label{sec:atmprofile}

\begin{figure}[h]
\includegraphics[width=\columnwidth]{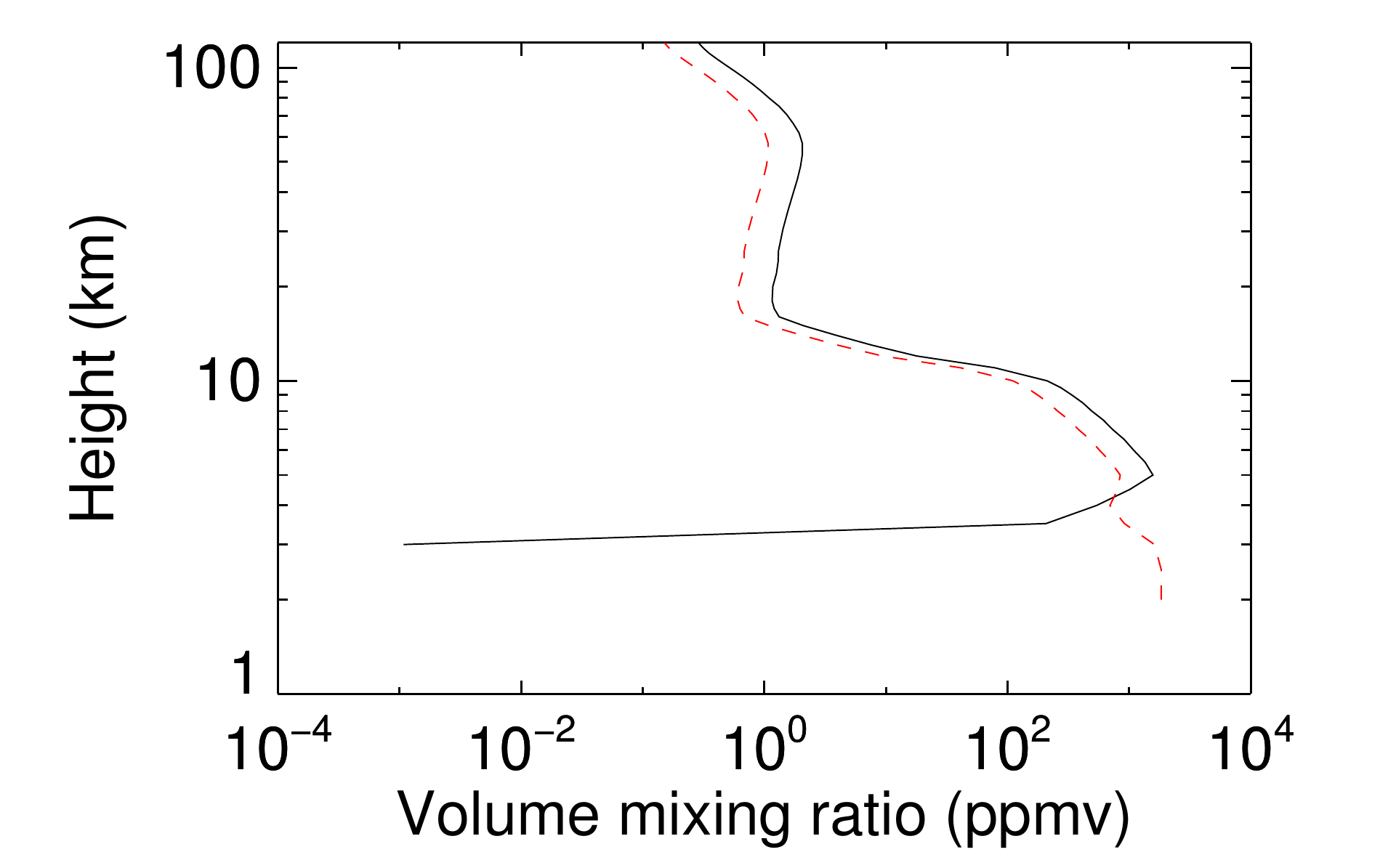}
\includegraphics[width=\columnwidth]{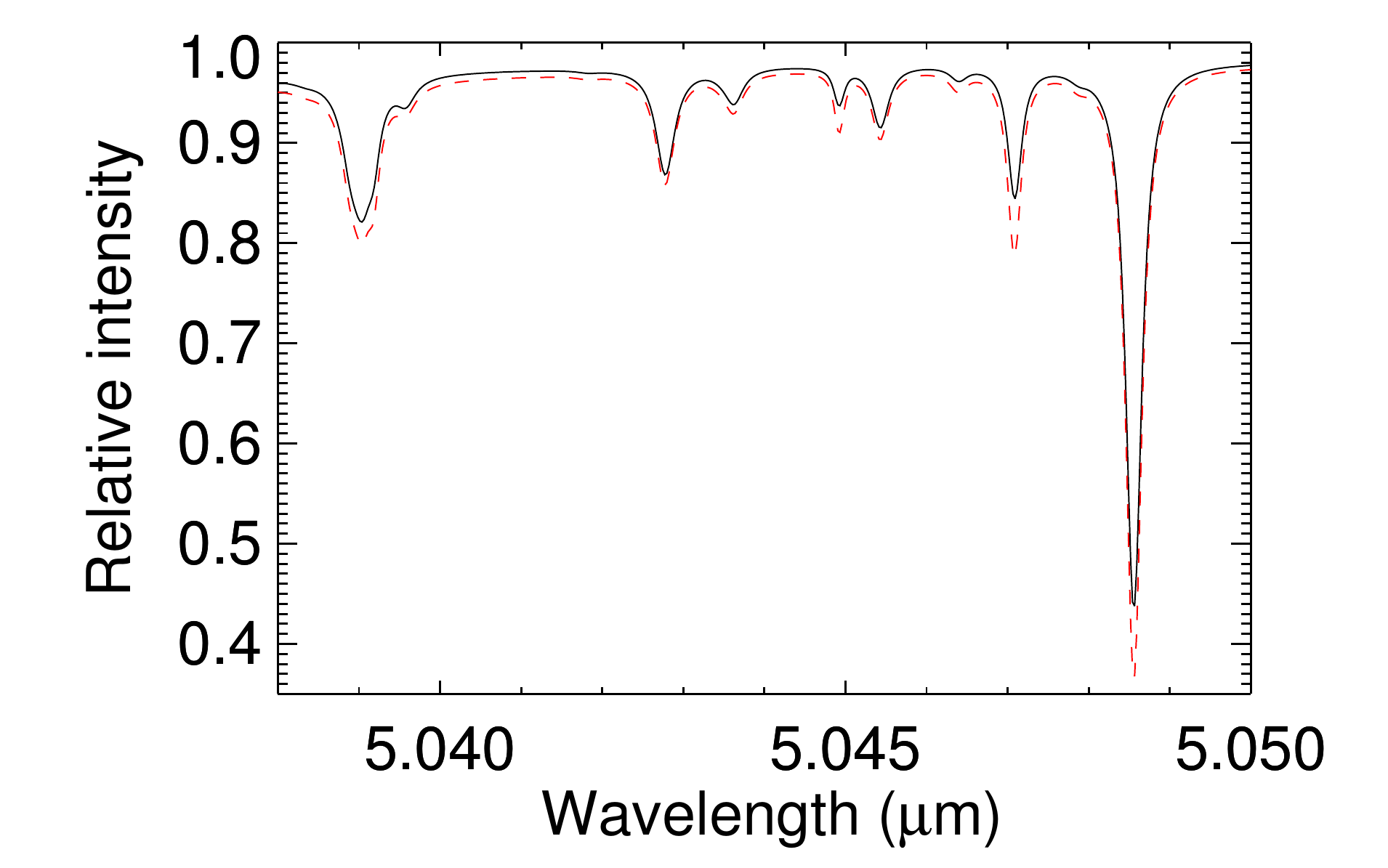}
\caption{\textit{Top:}  Two examples of  the distribution of  the water-vapour
  volume-mixing ratio as a function of altitude.  The black solid line
  corresponds  to a  volume-mixing  ratio of  0\,ppmv  up to  3.1\,km
  (Paranal altitude is 2.6\,km), while the red dashed line corresponds
  to a constant  relative humidity until 3.1\,km. The  total amount of
  precipitable water  vapour, 2\,mm,  is identical for  both profiles.
  \textit{Bottom:} sample of the corresponding transmission spectra at
  $R  = 100\,000$.  The  higher amount  of water  vapour in  the lower
  atmospheric layers above the  observatory leads to both deeper cores
  and stronger continuum absorption. }
\label{fig:pwv_artprof}
\end{figure}

An  atmospheric  profile  typically  describes  the  temperature $T$,
pressure $P$, and volume mixing ratio $x$ of several molecular species
as a function of altitude $h$ for a given location.  These parameters
have a  direct impact on  the shape and  strength of all  the telluric
lines. For \texttt{{\tt molecfit}} we  use a merged profile based on a
reference  atmospheric   profile,  modelled  3-D   data,  and  on-site
meteorological measurements.

The reference atmosphere profile is by default the equatorial
profile                  derived                 by                 J.
Remedios\footnote{\url{http://www.atm.ox.ac.uk/RFM/atm/}}. It contains
abundances of  several molecular  species up to  an altitude  of 120\,km,
divided  into 121 altitude  levels.  Owing  to the  low latitude  of Cerro
Paranal (Lat:  -24.6$^\circ$), the equatorial profile  is better suited
(Anu  Dudhia,  private communication)   than the  tropical  or  standard
  profiles, which are also available in the molecfit package. We note here that
  the molecular composition has  changed since the profile was created
  in 2001.  In  particular, the content of the  greenhouse-gas carbon-dioxide       content      has       increased       by      $\sim6\%$
\citepwmo{GHG_Bulletin_No.8_en.pdf}.   \texttt{Molecfit} allows  the user
to either  fit or manually adjust  the molecular content  to take this
increase into  account.  Alternatively, the  reference atmosphere used
by  \texttt{molecfit}  can  be  easily modified  to  another existing  or
manually created profile.

The    reference   atmosphere   is    then   merged    with   the modelled
GDAS\footnote{\texttt{http://www.ready.noaa.gov/gdas1.php}}   3-D   data
provided     by    the     National     Oceanic    and     Atmospheric
Administration\footnote{\texttt{https://ready.arl.noaa.gov/gdas1.php}}
(NOAA).   This  model  contains  time-dependent  information  on  the
temperature, pressure, relative humidity,  wind speed, and direction up
to an altitude  of $\sim26$\,km on a 3 h basis.   Owing to the geographical
1$^\circ\times1^\circ$ grid, we  created a set of profiles  by means of
an interpolation of  the four grid points closest  to Cerro Paranal to
obtain  local information. The  GDAS profile  derived for  a specific
observation  is then  by default  a  linear interpolation  of the  two
resulting  profiles closest  in  time to  the  observation.  The  {\tt
  molecfit} package contains all the available GDAS profiles for Cerro
Paranal since December 1, 2004 at 00:00 UT and the release date of the
package. In addition, ESO will  regularly maintain such a database for
Paranal. For  observations obtained later  or from another  site, {\tt
  molecfit} automatically retrieves the appropriate GDAS files.

It may still  be possible that no GDAS profiles  are available for the
time of  observations. In  the case of  Cerro Paranal,  \texttt{molecfit}
then   use  the   best   matching  two-month  average   \citep{2012A&A...543A..92N}.
Alternatively, an ASCII profile with the same format as a GDAS profile
can also be provided by the user.

A  further optional  refinement of  the final  profile is  achieved by
incorporating on-site measurements 
obtained by the Astronomical  Site Monitor (ASM)
as  provided in  the FITS  header or  through a  parameter  file.  The
influence  of  the measured  values  at  the  observatory altitude  is
gradually decreased up to a configurable upper mixing height,
set by default to $\sim5\,$km,  where, for Paranal, the wind direction
changes  by an  average of  $\sim90^\circ$ on  average, as
revealed by an analysis of the GDAS wind data. Thus, it can be assumed
that  at this  altitude the  influence  of the  local environment  (as
determined from the ASM data)  has diminished. More information on the
merging procedure can be found in the UM.

Water  vapour   is  the molecule showing the most variations 
because   its total amount (PWV) and its  altitude
distribution can  change significantly and on a  short time scale.  In
particular, the largest amount of PWV may not be located in the lowest
layers  of the  atmosphere.   
Figure~\ref{fig:pwv_artprof} illustrates
the importance of having a decent profile for water vapour.

The total amount  of PWV can optionally be set  if the value can
be   retrieved  independently  (for   example,  from    radiometer
\texttt{measurements}, see \citetads{2012SPIE.8446E..3NK}).  In  this  case, the  merged  profile
composed  by the  reference atmosphere,  the GDAS  data, and  the local
meteo  information will be  scaled to the requested PWV. 
Alternatively, the atmospheric profile provided by the radiometer
-- soon retrievable from the ESO archive -- can be used directly. 

\begin{figure}[h]
\includegraphics[width=\columnwidth]{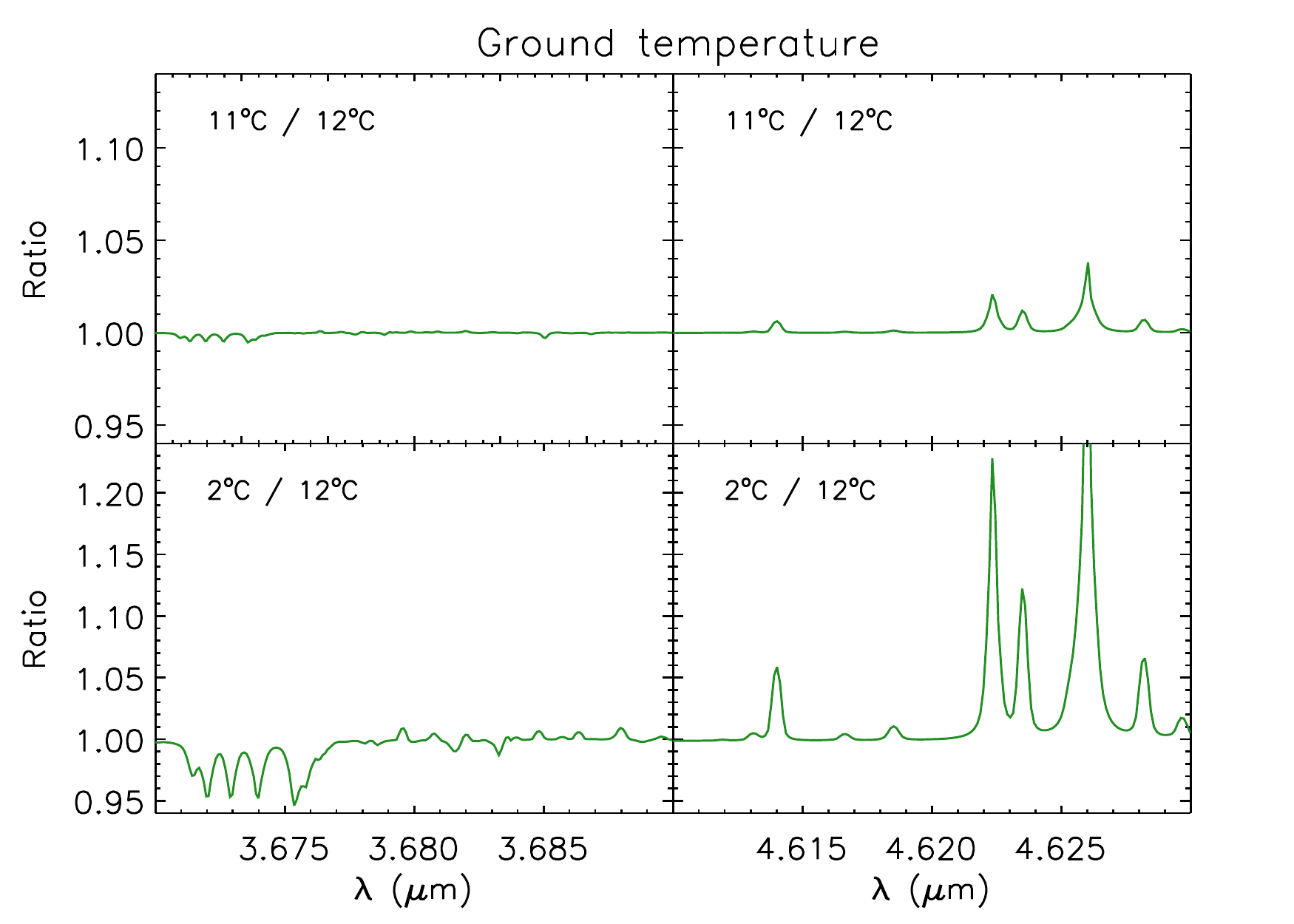}
\includegraphics[width=\columnwidth]{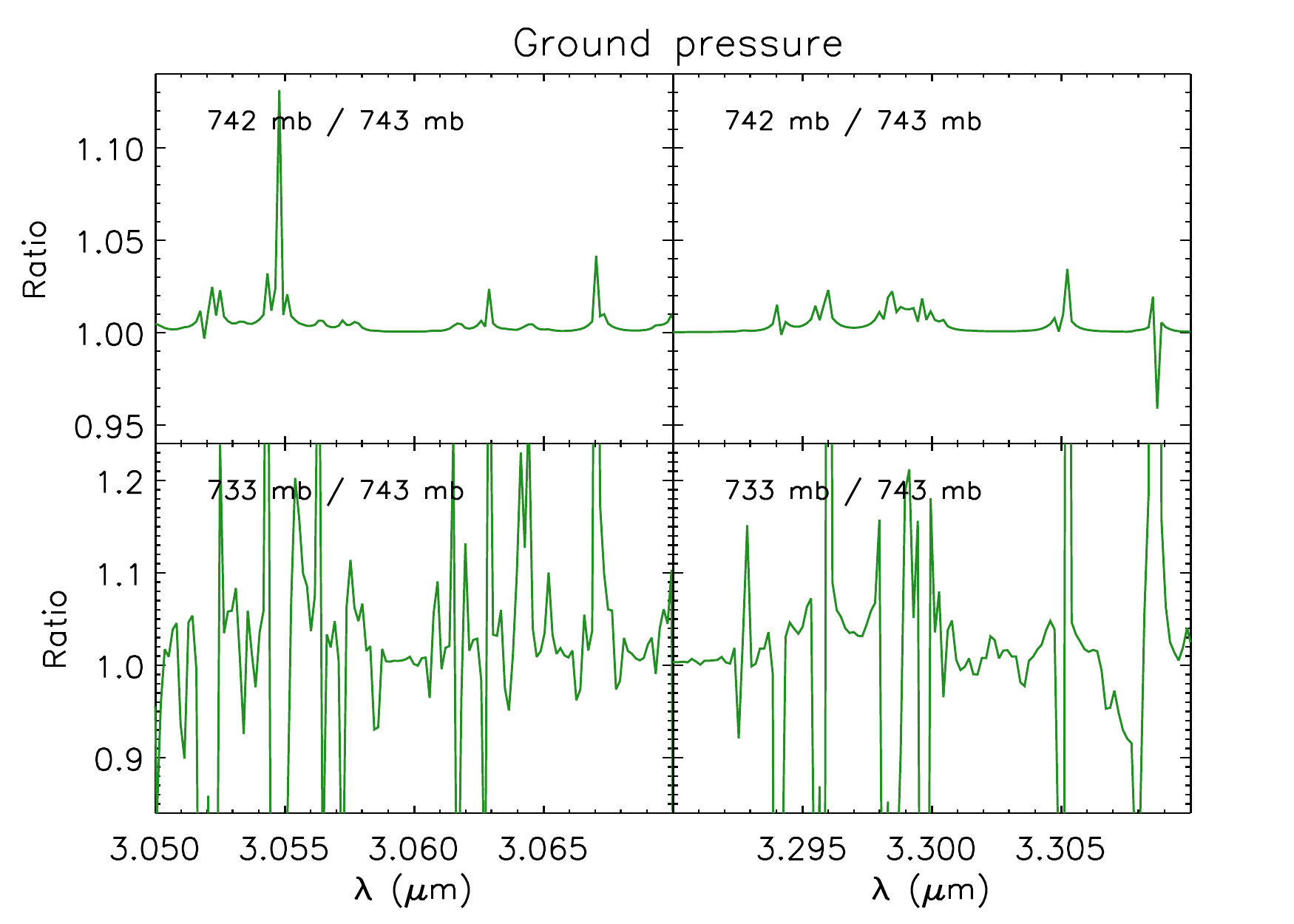}
\caption{ Effect  of ground temperature  or pressure variation  on the
  H$_2$O telluric  spectrum for  two different spectral  ranges.  Each
  graph  shows the ratio  between telluric  spectra computed  with the
  indicated   ground  temperature   or  pressure.    \textit{Top: }
  variation  of 1\,$\degr$C  and 10\,$\degr$C.   The region
  close to 3.68\,$\mu$m shows opposite behaviours. \textit{Bottom:}
  variation of 1~mb and 10~mb.}
\label{fig:gtemppres}
\end{figure}

Figures~\ref{fig:gtemppres}  illustrates  the  effect  of the
variation of the ground temperature and pressure on the
H$_2$O  telluric spectrum.   Depending on  line  parameters, different
lines  behave differently with temperature or pressure.

\subsection{Radiative transfer code}\label{sec:lblrtm}

The  resulting merged  atmospheric profile  is  used as  input in  the
radiative  transfer  code  LNFL/LBLRTM  \citep{CLO05}. The
  version used at the time of the first release of the
  \texttt{molecfit} package is the latest
version at that time,  V12.2.   This code  package  is  widely  used in  atmospheric
sciences  and  is  publicly  available.  It  calculates  radiance  and
transmission  spectra  with a spectral resolving power  of  about  four million.  More
information       can       be       found      at       the       web
site\footnote{\url{http://rtweb.aer.com/lblrtm_frame.html}}.

\subsection{Database of molecular parameters}
\label{sec:database}

In addition  to the  profile, the radiative  transfer code  requires a
line database as input.  The first release of 
  \texttt{molecfit} uses the database \texttt{aer\_v\_3.2}, which
is  delivered with the  LNFL/LBLRTM code  package: it  is based  on an
updated version of the HITRAN 2008 database \citep{ROT09} and contains
information on  more than 2.7  million spectral lines of  42 molecular
species  (see  Table   \ref{tab:molecs}).   The  \texttt{ molecfit}  line
database  can be  easily updated  when a  new version becomes available.

Only  molecules  for which  an  atmospheric  volume mixing ratio  is
available in the standard atmosphere profile \texttt{provided with} \texttt{molecfit} can be
used   by   default. These   molecules   are   identified  in   Table
\ref{tab:molecs}.

\begin{table}[!Ht]
\begin{center}
  \caption{List  of  molecules  as  provided  by the \texttt{ aer}  line
    parameter  database.} 
\vspace{6pt}
\centering
\begin{tabular}{r | c | c | c | c}     
\hline\hline
     \#
   & Molecule           
   & Chemical name           
   & In standard
   & GUI\\
   & 
   & 
   & profile?
   &\\
\hline
 1 & H$_2$O             & Water               & X & f \\
 2 & CO$_2$             & Carbon dioxide      & X & f \\
 3 & O$_3$              & Ozone               & X & f \\
 4 & N$_2$O             & Nitrous oxide       & X & f \\
 5 & CO                 & Carbon monoxide     & X & f \\
 6 & CH$_4$             & Methane             & X & f \\
 7 & O$_2$              & Oxygen              & X & f \\
 8 & NO                 & Nitric oxide        & X & c \\
 9 & SO$_2$             & Sulfur dioxide      & X & c \\
10 & NO$_2$             & Nitrogen dioxide    & X & c \\
11 & NH$_3$             & Ammonia             & X & c \\
12 & HNO$_3$            & Nitric acid         & X & c \\
13 & OH                 & Hydroxyl            &   &   \\
14 & HF                 & Hydrogen fluoride   &   &   \\
15 & HCl                & Hydrogen chloride   &   &   \\
16 & HBr                & Hydrobromic acid    &   &   \\
17 & HI                 & Hydrogen iodide     &   &   \\
18 & ClO                & Chlorine monoxide   & X & c \\
19 & OCS                & Carbonyl sulfide    & X & c \\
20 & H$_2$CO            & Formaldehyde        &   &   \\
21 & HOCl               & Hypochlorous acid   & X & c \\
22 & N$_2$              & Nitrogen            & X & c \\
23 & HCN                & Hydrogen cyanide    & X & c \\
24 & CH$_3$Cl           & Chloromethane       &   &   \\
25 & H$_2$O$_2$         & Hydrogen peroxide   & X & c \\
26 & C$_2$H$_2$         & Acetylene           & X & c \\
27 & C$_2$H$_6$         & Ethane              & X & c \\
28 & PH$_3$             & Phosphine           &   &   \\
29 & COF$_2$            & Carbonyl fluoride   & X & c \\
30 & SF$_6$             & Sulfur hexafluoride & X & c \\
31 & H$_2$S             & Hydrogen sulfide    &   &   \\
32 & HCOOH              & Formic acid         &   &   \\
33 & HO$_2$             & Hydroperoxyl        &   &   \\
34 & O                  & Oxygen              &   &   \\
35 & ClONO$_2$          & Chlorine nitrate    & X & c \\
36 & NO+                & Nitrosonium         &   &   \\
37 & HOBr               & Hypobromous acid    &   &   \\
38 & C$_2$H$_4$         & Ethylene            &   &   \\
39 & CH$_3$OH           & Methanol            &   &   \\
40 & BrO                & Bromine Monoxide    &   &   \\
41 & C$_3$H$_8$         & Propane             &   &   \\
42 & C$_2$N$_2$         & Cyanogen            &   &   \\
\hline
\label{tab:molecs}
\end{tabular}
\tablefoot{
  Col. 1:  reference
    number of the molecule as defined in the HITRAN 2008 database; Col. 2:  its chemical  formula,  followed  by its  chemical
    name. Col. 3 indicates if an atmospheric profile of the
    molecular  volume mixing ratio is  available in  the  equatorial standard profile
    as included in\texttt{ molecfit}.  Last column:
    functionality   available   through   the  GUI:   \textit{f}   the
    corresponding  column  density   can  be  fitted,  \textit{c}  the
    corresponding transmission  spectrum is only calculated  using the user
    provided  abundance, relative to  the one  fixed in  the reference
    atmosphere.}
\end{center}
\end{table}


\subsection{Fitting algorithm}
\label{sec:fittinglagorithm}

\texttt{Molecfit} makes  use of  the C  version of  the  least-squares fitting
library MPFIT \citepads{2009ASPC..411..251M} based on the FORTRAN fitting routine
MINPACK-1 \citepdoi{BFb0067700}. It uses the Levenberg-Marquardt
technique to solve the least-squares problem leading to a fast
convergence even for the complicated functions involved in
\texttt{molecfit}, while being numerically robust and self-contained
and allowing upper and lower limits or
parameters tied to each other that can be individually set.

\subsection{Inclusion and exclusion regions}
\label{sec:IncludingAndExcludingRegions}

A spectrum shows a number of features that can be
characterised in the following way:
\begin{itemize}
\item intrinsic features from the science object,
\item telluric features,
\item intrinsic features from the instrument caused either by the
  spectrograph or the detector.
\end{itemize}
Intrinsic features either from the science object or from the instrument can
interfere with the fitting process. 

Therefore, \texttt{molecfit} allows one to:
\begin{itemize}
\item select inclusion regions, also called \emph{\emph{fitting ranges}}, whose main
  purposes are to allow\texttt{ molecfit} to determine the column density
  of the relevant molecules and  the line spread function
  (or kernel). Such regions must therefore show a good enough
    continuum level for the fitting accuracy and include telluric absorption lines
    for determining the line spread function. One or several
    such regions must be defined so that at least some telluric lines from each
    molecule to be corrected are covered.
\item select exclusion regions  within the inclusion regions that are
  affected  by the  presence of  intrinsic features  from  the science
  objects;
\item mask pixels  in the inclusion regions that  are affected purely
  by instrumental or detector effects.
\end{itemize}
Details can be found in the UM. 

\subsection{Continuum contribution}
\label{sec:continuum}
The model spectrum, $F_\mathrm{out}$, is scaled by a polynomial of degree $n_\mathrm{c}$:
\begin{equation}
F_\mathrm{out}(\lambda) =
F_\mathrm{in}(\lambda) \, \sum_{i = 0}^{n_\mathrm{c}} a_i \lambda^i.
\end{equation}

For $a_0  = 1$  and all other  $a_i =  0$, the model  spectrum remains
unchanged.  This   is  the  default  configuration   for  the  initial
coefficients. 
The continuum correction is carried
out independently for each fitting range.
A fitting range (or the full spectrum) is split further
if it is distributed over more than one chip.

If the input spectrum is an emission line spectrum, an optional flux
conversion can be carried out in order to match the unit of the
synthetic spectrum. Further details are available in the UM.

\subsection{Telescope background}
\label{sec:backgrounds}

For sky emission modelling, the  telescope background is assumed to be
a grey body (black body times emissivity). The parameters of this grey
body  correspond  to the  telescope  main  mirror  temperature and  an
effective emissivity of the telescope and instrument.

\subsection{Wavelength calibration}
\label{sec:wavelength}

The  science  spectra  to  analyse  are  expected  to  have  an accurate
wavelength  calibration.  However, small errors can affect
the determination of the line spread function and column densities negatively. 
Therefore in  each  inclusion  region,  the synthetic spectrum being fitted can
be adjusted to match the input spectrum.
A Chebyshev polynomial of degree $n_\mathrm{w}$ is used for this purpose:
\begin{equation}
\label{eq:wavelength}
\lambda = \sum_{i = 0}^{n_\mathrm{w}} b_i t_i,
\end{equation}
where
\begin{equation}
  t_i = \left\{ 
    \begin{array}{ll} 
      1 & \textrm{for\ } i = 0 \\
\lambda & \textrm{for\ } i = 1 \\
2 \, \lambda^\prime \, t_{i-1} - t_{i-2} & \textrm{for\ } i \ge 2
\end{array} \right.
\end{equation}
where the wavelength range  
is normalised so that $\lambda^\prime$ ranges from $-1$ to $1$
over the entire spectral range of the spectrum or, in the case of
multi-chip instruments, over the range covered by an individual chip. 

The  conversion of the  wavelength grid to a  fixed interval
results in coefficients $b_i$  independent of the wavelength range and
step size of  the input spectrum. For  $b_1 = 1$ and all  other $b_i =
0$,  the  model  spectrum  remains  unchanged.  This  is  the  default
configuration for the initial coefficients. 

On the other hand, in some cases, one wishes to use the telluric lines
as references to improve an unsatisfactory initial wavelength
calibration.  This is  the case for CRIRES in  particular.  For such a
purpose,  \texttt{molecfit}  actually   outputs  the  inverse  of  the
wavelength calibration  described in Eq.   \ref{eq:wavelength}, in the
sense that the input spectrum  is then adjusted to match the synthetic
one. Such wavelength  calibration is only meaningful when
the inclusion range  covers the whole spectrum and  is sampled well by
telluric lines.

\subsection{Line spread function}
\label{sec:kernel}

The model spectrum is convolved with up to three different profiles to
determine the shape of the line spread function. They are:
\begin{enumerate}
\item a simple boxcar
\begin{equation}
F_\mathrm{box}(\lambda) = \left\{ \begin{array}{ll}
1 & \textrm{for\ } -w_\mathrm{box}/2 \le \lambda \le w_\mathrm{box}/2 \\
0 & \textrm{for\ } \lambda < -w_\mathrm{box}/2 \ \mathrm{or} \lambda > w_\mathrm{box}/2
\end{array} \right.
,\end{equation}
which is adapted to the pixel scale and normalised to an integral of 1.
This kernel is particularly useful for objects fully covering the
entrance slit; and\\

\item a Gaussian
\begin{equation}
F_\mathrm{gauss}(\lambda) = \frac{1}{\sigma \sqrt{2 \pi}}
\exp{\Bigg(-\frac{\lambda^2}{2 \sigma^2}\Bigg)}
\end{equation}
centred on 0, where
\begin{equation}
\sigma = \frac{w_\mathrm{gauss}}{2 \sqrt{2 \ln{2}}}.
\end{equation}
The full width  at half maximum (FWHM), $w_\mathrm{gauss}$,  is given in
pixels. \\

\item  a Lorentzian
\begin{equation}
F_\mathrm{lorentz}(\lambda) = \frac{1}{\pi} \,
\frac{\lambda}{\lambda^2 + (w_\mathrm{lorentz}/2)^2}
\end{equation}
centred on 0, where $w_\mathrm{lorentz}$ is the FWHM in pixels. Compared to a
Gaussian, the Lorentzian approaches the 0-level flux significantly more slowly.
\end{enumerate}

A kernel consisting of all three components is usually too complex 
to be constrained by a fit. To avoid unrealistic best-fit FWHM, the user should 
reduce the number of degrees of freedom by fixing individual fit
components. A FWHM of zero pixel corresponds to a unity convolution;
in other words, it does not change the input spectrum.
   
Alternatively, the  user has the possibility of choosing
a  single  Voigt  profile  kernel directly, which  is then calculated  by  an
approximate formula that takes the  FWHM of Gaussian and Lorentzian as
inputs.
The  user can  also  select  a kernel  whose  size in  pixels
linearly   increases  with   wavelength.   It   is  suitable   for  an
instrumental  setup  with   nearly  constant  resolving  power,  fixed
wavelength   step  size,  and   a  wide   wavelength  range,   such  as
cross-dispersed echelle spectrographs.

Finally, a user-defined kernel can be provided in the form of an ASCII
table   of  fixed   kernel   elements.  This   option  overrules   the
parameterised kernel discussed above.
An inappropriate  kernel can cause deviations in  the determination of
molecular column densities of more than 10\%.

\subsection{Molecular volume-mixing ratio}
\label{sec:columndensity}

The integrated  volume mixing ratio,  $X(h_0)$, of a molecule  for a
column of air  from the observer at an  altitude $h_0$ to the
top of atmosphere can be derived by
\begin{equation}
X(h_0) = \frac{\int_{h_0}^{\infty}{\frac{x(h)\,P(h)}{T(h)}\,\mathrm{d}h}}
{\int_{h_0}^{\infty}{\frac{P(h)}{T(h)}\,\mathrm{d}h}}
\end{equation}
if the mixing ratio (or mole fraction) $x$, air pressure $P$, and temperature $T$
are given depending on altitude $h$ by an atmospheric profile. Volume-mixing
ratios are usually given in part per million in volume (ppmv).

For $\mathrm{H_2O}$, it is also common to provide the column height of
PWV in mm (also referred to as amount of PWV). It can be calculated by
\begin{equation}
\mathrm{PWV} = \frac{m_\mathrm{mol, H_2O}}{\rho_\mathrm{H_2O}\,R}
\int_{h_0}^{\infty}{\frac{x_\mathrm{H_2O}(h)\,P(h)}{T(h)}\,\mathrm{d}h},
\end{equation}
where the mole mass of water $m_\mathrm{mol,\,H_2O} = 0.0182$\,kg, the density
of liquid water $\rho_\mathrm{H_2O} \approx 10^3\,\mathrm{kg\,m}^{-3}$, and the
gas constant $R = 8.31446\,\mathrm{J\,mol^{-1}\,K^{-1}}$. The required units of
$x$, $P$, $T$, and $h$ are ppmv, Pa, K, and km, respectively. {\tt
  Molecfit} outputs the column height for other gases using a similar formula.

Since the GDAS profiles and the ESO MeteoMonitor data provide an $\mathrm{H_2O}$
volume-mixing ratio as relative humidity $RH$ in percent, it has to be converted to
$x$. For this purpose, we have implemented the approximations for vapour
pressures of ice and supercooled water as function of temperature as described
by \citetdoi{10.1256/qj.04.94}. The minimum of
both then corresponds to the saturated vapour pressure $P_\mathrm{sat}$, which
is used to estimate the altitude-dependent $x$ in ppmv by
\begin{equation}
x(h) = 10^6 ~ \frac{P_\mathrm{sat}(T(h))}{P(h)}\,\frac{RH(h)}{100}.
\end{equation}

\texttt{Molecfit} allows the user to adjust this quantity based on the observed telluric
lines, and we note that the GUI only allows one to fit the column density for
the molecules indicated in the last  column of Table \ref{tab:molecs}.
This adjustment is done by multiplying the overall atmospheric
profile by a single constant. For most molecules, the impact is
actually limited to the lowest atmospheric layers.

\subsection{Supported instruments}
\label{sec:supportedinstruments}

One-dimensional (1-D) spectra of  any visible, near-, or mid-infrared
instrument could in  principle be used by\texttt{ molecfit}.  Details on
the  required   information  can  be  obtained  in   the  User  Manual
\citep{NOL13}.  Although the  tool  has been  tailored to  ESO-Paranal
instruments --  in particular,  the GDAS profiles  are available  as a
tarball file only  for this site -- there  are no specific limitations
either  in   the terms  of  format   or  in geographical  location   for  a
ground-based observatory.

\subsection{Inputs and outputs}
\label{sec:outputs}

\texttt{ Molecfit} accepts  1-D spectra  provided as  ASCII  tables, FITS
tables, and   images. The FITS tables can  have several extensions
corresponding to different chips  (e.g., CRIRES). Required keywords are
either  taken from  the FITS  header  or directly  from the  parameter
file. Finally, ASCII or FITS files for inclusion and exclusion regions
can be provided.

\texttt{ Molecfit} fits the  synthetic spectrum  in the  inclusion regions
alone. These regions can cover just  a small fraction of the
wavelength  range or all of it.   It  returns  the  best-fit model  only  for  the
inclusion regions in both an ASCII  and a FITS table, and the best-fit
parameters  and molecular column  densities (including  the PWV)  in a
results ASCII file.

The \texttt{molecfit} package then outputs  a transmission spectrum and the
division of the  input spectrum by the transmission  spectrum over
the full spectral range of the input  spectrum. 
Apart from results tables in ASCII and FITS format, the corrected spectrum is
written into a file in the same format as the input spectrum (see the UM for
more details)

In addition, \texttt{molecfit}  can provide a new  wavelength calibration
based  on  the telluric  lines as described in Sect.~\ref{sec:wavelength}.
This two-step approach greatly improves the speed of the overall process. 

\subsection{Expert mode}
\label{sec:expert}

In  most cases \texttt{molecfit}  is able  to determine  the different
user-selected   parameters    automatically   with   a    minimum   of
configuration.  However, in  a few  cases, mostly  for CRIRES,  a much
greater flexibility is needed.
An expert mode is therefore also  provided.  It allows the user to fit
the spectra on each  inclusion range individually and provides him/her
the access  to the coefficients  of the polynomials for  the continuum
and  wavelength correction.   Moreover,  the parameter  file with  the
best-fit  values for  all  parameters is  saved  and can  be used  for
another iteration of molecfit.

\subsection{Fitting sky emission spectra}
\label{sec:fittingskyemission}

The  tool  is able  to  fit sky  emission  spectra  for molecules  and
physical processes  handled by the radiative  transfer code, with the 
exclusion of lines produced by chemiluminescence (such as OH, see,
e.g.,  \citetads{2012A&A...543A..92N}). In other words \texttt{molecfit}
can in principle determine the line spread function and column densities
of the relevant molecules and therefore derive the transmission
spectrum of the atmosphere based  the sky  emission spectrum if
the spectral range  is located in the thermal  infrared.

\section{Impact on observing strategy}
\label{sec:usage}

Although  powerful,\texttt{ molecfit}  is not  necessarily  suited  to
telluric corrections in all cases. In this section, we briefly discuss
a few situations to consider while deciding on the details of the
observing strategy for the absorption telluric correction.

\subsection{Main cases}
\label{sec:maincases}
Two main cases for the use of \texttt{molecfit} can be considered, depending
on whether the wavelength calibration  of the spectrum is reliable and
accurate (root-mean-squared residuals  $\approx 0.1$ pixel or better) or not.

In the  case of accurate wavelength calibration,  the main requirement
is that the spectrum to be  corrected has at least one spectral region
that  includes medium-strength  molecular lines  of the  most variable
relevant  species and  that these  lines are  isolated enough  so that
their column density and the line spread function can be determined.

If  the  wavelength  calibration  is  not reliable  but  needs  to  be
corrected by\texttt{  molecfit}   based  on  the   telluric  lines
themselves,  a --  possibly  delicate --  selection  of inclusion  and
exclusion regions must  be carried out, such that  only telluric lines
appear in the inclusion regions and that any intrinsic feature of the
object is masked  out in the exclusion regions.  In case the continuum
needs to be refined by  \texttt{molecfit}, the fitting regions need to
present enough coverage to be  able to determine either an overall
continuum or local continua.

\subsection{Simple continuum shape and small number of intrinsic features}
\label{sec:simple}

The easiest spectra that can be  modelled by the tool are the ones for
objects  with a  well-defined and  simple continuum  and  with limited
confusion   between  telluric   and  intrinsic   features.    In  such
conditions, the tool can properly  model the continuum spectrum of the
object,  and  the parameters  of  the  telluric  features can  be  well
constrained.

\subsection{Large number of intrinsic features}
\label{sec:numerousintrinsic}

If the number of intrinsic features is large in the spectral domain of
interest, several options are possible.  The choice of the best option
depends  on the line  crowding: a  first constraint  is the  number of
points that can be used to determine the spectral continuum; a second
constraint is the  number of spectral ranges where  telluric lines can
be used to determine the  column density of the molecules contributing
in  this spectral  domain.   An additional  constraint  arises if  the
wavelength calibration is not reliable.

Therefore a possible strategy is first to observe a telluric star with
the same setup --  or retrieve corresponding data from  the archive --
and  use \texttt{ molecfit} on  its reduced  spectrum to  determine which
molecules  are  relevant,  the  wavelength calibration  or  continuum
determination.   Then, \texttt{molecfit}  can be  applied to  the science
target  using  the   derived  wavelength  calibration  (if  relevant),
changing  the  airmass, or  fitting  the  continuum  and/or the  column
densities of the molecules involved.  An example of such a strategy is
shown   below  in   Sect.~\ref{sec:crires}  and   is   illustrated  in
particular in Fig.~\ref{fig:crires_4889p5_e}.     Another   possible
strategy  is  to ensure  that  the  spectral  setup includes  spectral
regions containing representative telluric lines but free of
intrinsic lines.

\subsection{Emission line spectra}
\label{sec:nocontinuum}

The telluric transmission spectrum cannot be fitted on the spectrum of
an object  showing little  or no continuum,  such as a  faint cometary
spectrum or  an emission  line object.  A  suggested strategy  in this
case is  either to  retrieve a  spectrum of a  telluric star  from the
archive or observe a telluric star in the spectral set-up of interest.
Its spectrum can be modelled by \texttt{molecfit}.  The model can then be
modified easily  by changing the  airmass such that it  corresponds to
the airmass of the science target.  If the water vapour at the site is
monitored, the model can also  be modified accordingly.  The number of
times  telluric  stars  need  to  be  observed  is  therefore  reduced
significantly.

\section{Limitations}
\label{sec:limitations}

In this section, we briefly discuss the various factors that limit
the accuracy of the \texttt{molecfit}  correction of
telluric spectra. Such factors can be external to \texttt{molecfit},
and are related to the accuracy with which instrumental  parameters
are taken into account or are internal to the method itself.

\subsection{External}
\label{sec:externallimitations}

\paragraph{Molecular parameters' accuracy and completeness: }
A first limitation arises owing  to the completeness of the incorporated
line  database  and  on   the  accuracy  of  the  contained  molecular
parameters.  Currently, the AER  line database delivered with
the LBLRTM  code is  included by default  with \texttt{molecfit}.   It is
usually  more  frequently  updated  than the  underlying  main  
HITRAN\footnote{\url{http://www.cfa.harvard.edu/HITRAN/}}   and   is
therefore  recommended.  However,  although \texttt{ molecfit}  \emph{\textit{\emph{a
  priori}}  }allows  one  to  use  the original  HITRAN  database,  this
functionality  is subject  to the stability of the
database structure.

\paragraph{Molecules in atmospheric profile:}
\texttt{Molecfit} only  offers the possibility to fit  the overall volume
mixing ratio  along the line of sight  in the atmosphere:  the overall
shape of  the profile is not  adjusted. If the  actual profile differs
from the  modelled one in  some atmospheric layers,  systematic errors
can occur.   In particular such a situation  can be caused by  the specific, temporary
volume-mixing ratios  of a molecule and  is  relatively frequent for
water     vapour.  Figure~\ref{fig:pwv_artprof} illustrates the possible 
impact in  an    extreme     case.

In particular, the  accuracy of  the transmission  spectrum derived
from sky emission as described in Sect. \ref{sec:fittingskyemission}
strongly depends on the precision with which the retrieved atmospheric
profile  actually represents  the  true one.

\paragraph{Atmospheric profile stability: }
Any  correction of  telluric absorption  requires a  relatively stable
atmosphere.  The arrival of an atmospheric front is a clear case where
the atmospheric  profile changes suddenly.  Even if the  GDAS profiles
would show  such a front,  its precise timing and  characteristics are
usually  uncertain  and  surely   depend  on  the  pointing  direction.
Fortunately,  few optical or  infrared astronomical  observations are
executed in such conditions.

\paragraph{Radiative transfer code accuracy:}
The  reader is  referred to  the  publication list  available on  the
LBLRTM                                                              web
page\footnote{In particular, \url{http://rtweb.aer.com/lblrtm_ref_pub.html}}       for
references regarding the validation of the radiative transfer code.
In particular, it  is worth repeating  here  that ``the  algorithmic
accuracy of  LBLRTM is approximately  0.5\% and the  errors associated
with the computational procedures are  of the order of five times less
than those  associated with the  line parameters so that  the limiting
error is that attributable to the line parameters and the line shape''.

\subsection{Instrumental}
\label{sec:instrumentallimitation}

\paragraph{Grating scattering: }
Internal  diffusion strongly  limits  the achievable  accuracy of  the
modelling.   Villanueva et  al. (2009,  internal  communication) found
that the  CRIRES line  spread function is  best reproduced by  a Voigt
profile that  needs to  be calculated over  $\approx 1\,000$  pixels  to   account  for  the  grating  scattered   light  within  the
instrument.   As a  consequence, it  often  appears that  the core  of
heavily saturated absorption lines  does not reach zero intensity. The
amount of  flux in  the core actually  depends on the  spectral energy
distribution of the light reaching  the grating and cannot be modelled
accurately  by\texttt{ molecfit},  because\texttt{ molecfit}  only applies  the
convolution of  the line spread function to  the transmission spectrum
and not to  the science target one.  In  addition, later investigation
(Uttenthaler 2010,  private communication) indicates  that the spatial
profile of  the grating scattered light  is different from  the one of
the direct  light. 

The  observed  spectrum  therefore  combines different  effects  that
cannot be  easily modelled by  \texttt{ molecfit}.   Therefore they limit
the  accuracy of  the  correction  such that  residuals  can reach  an
estimated 1\%  of the continuum  depending on the significance  of the
grating scattering.

\paragraph{Instrumental background: }
When modelling sky emission spectra, \texttt{molecfit} 
uses a background defined by a grey body at the temperature
of the main mirror of the telescope. It assumes that the
thermal background from the instrument is negligible.

\paragraph{Instrumental response curve shape:}
\texttt{Molecfit} fits  a low-order polynomial  to
the  observed continuum of  the input  (object spectrum) for each  inclusion region.   However, it
does not make a difference  between the  intrinsic continuum  shape  and the
effect  of the  response curve  of  the instrument.  Good results  are
obtained  if  the  observed  continuum  of  each  inclusion  region  can
be modelled well by such polynomials. Alternatively, the user may
provide a normalised spectrum as input.

\paragraph{Line spread function: }  The line spread function should behave
well   over  the wavelength  range  of  interest, either  being
constant or having a width that is linearly dependent on wavelength, as expected
in cross-dispersed spectrographs.

When  the transmission spectrum  is derived from  the sky
emission  spectrum  (Sect.  \ref{sec:fittingskyemission}), one  should
note  that the  parameters  of  the line  spread  function are  likely to be
different from the  one of the science target given  that the image of
the star does not uniformly illuminate the entrance slit.

\subsection{Internal}
\label{sec:internallimitations}

\paragraph{Good internal guess values:}
\label{sec:goodguessvalues}
By essence of the Levenberg-Marquardt fitting approach, \texttt{molecfit}
needs and is  sensitive to the initial guess  solution. In particular,
\texttt{molecfit}  is designed to  refine the wavelength  calibration and
continuum determination, but it is not designed to measure these values
from  an uncalibrated  spectrum.  Adequate  initial  estimates improve
the robustness  of  the fit  and  the speed  of the  convergence
process. Knowledge  of the line spread function  and spectral resolving
power are  also important,  although the latter  can be adjusted  by {\tt
  molecfit}.

\paragraph{Finite pixel size:}
\texttt{Molecfit}  results are  best when the  spectral resolution  is at
least Nyquist sampled  by the instrument. In any  case, \texttt{molecfit}
takes the finite  pixel size of the spectrum into account: \texttt{molecfit}
estimates the highest pixel resolution in a spectrum by evaluating the
wavelength grid.  An oversampling factor of  5 is then  applied on the
input to the radiative transfer code.

\begin{figure*}[ht]
  \sidecaption
  \includegraphics[width=12cm]{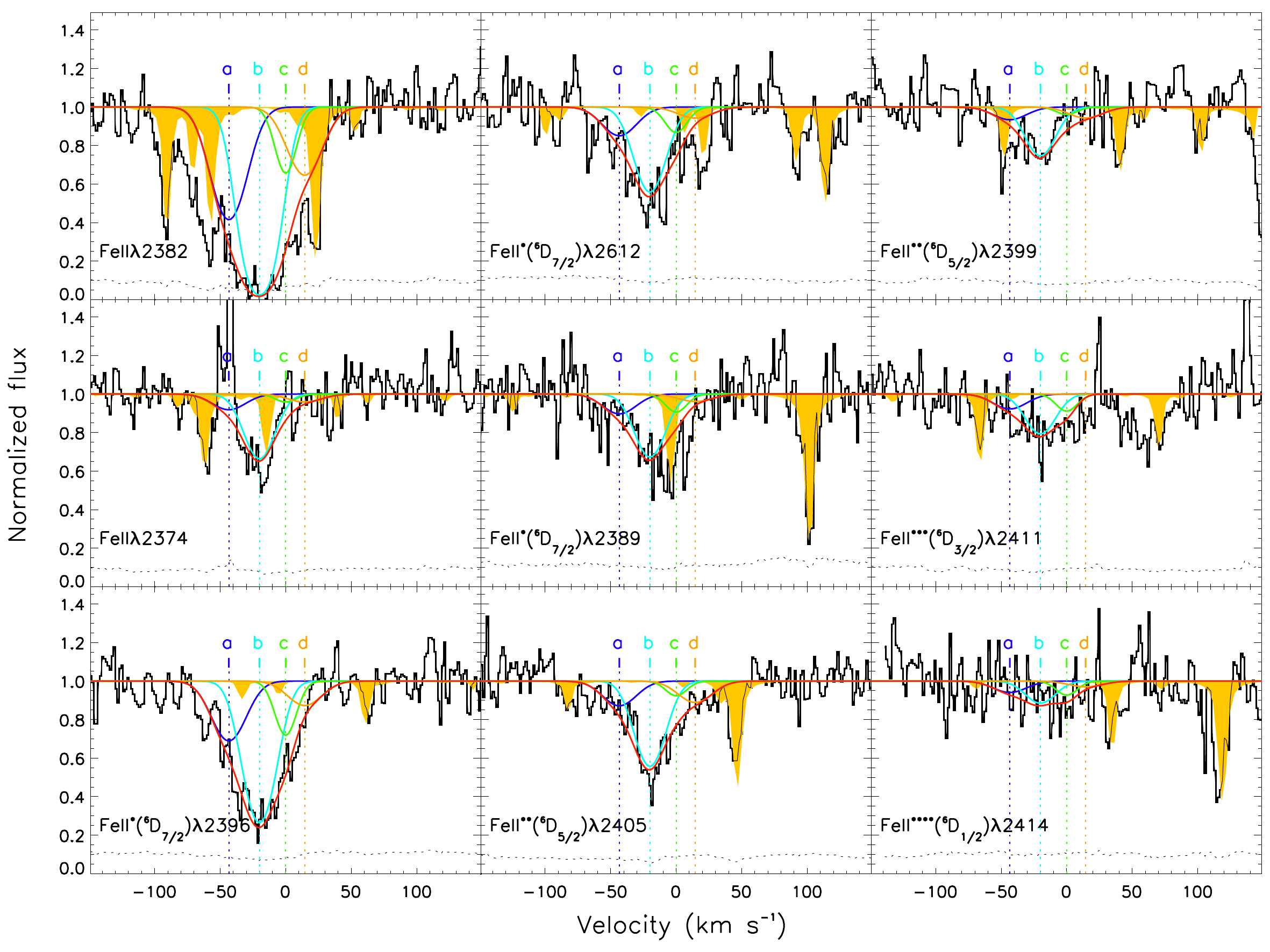}
  \caption{$\mathrm{H}_2\mathrm{O}$    telluric    absorption    lines
    affecting absorption  lines associated with the $z  = 2.42743$ DLA
    in  the UVES spectrum  of  \object{GRB080310}. The  telluric  lines  -
    modelled by \texttt{ molecfit} - are shaded in  yellow; the models for
    the different components ($a,b,c,d$)  of the absorption system are
    represented  with different  colours (blue,  cyan,  green, orange),
    while     the     combined     model     is    in     red.     See
    \citetads{2012A&A...545A..64D} for details.}
\label{fig:grb080310}
\end{figure*}

\section{Examples}
\label{sec:examples}

In the  following, we illustrate the  versatility of \texttt{molecfit}
through its  application to  data sets obtained  with various  ESO VLT
instruments covering a range of spectral resolving powers and
  wavelength domains.

\subsection{UVES}

The  prototype version  of \texttt{ molecfit} was  manually
  adjusted to  determine the
location and  strength of telluric absorption lines  in  VLT UV--Visual
\'Echelle  Spectrograph  (UVES,  \citeads{2000SPIE.4008..534D})  Rapid
Response  Mode or  Target-of-Opportunity  programmes of  $\gamma$-ray
bursts (UVES is a  high-resolution cross-dispersed spectrograph in the
visible):  indeed,  some  $\mathrm{H}_2\mathrm{O}$ lines  mimic  lines
associated  with high-redshift  damped Ly-$\alpha$  (DLA)  lines.  For
example,      in      the      spectrum     of      \object{GRB050730}
\citepads{2009A&A...506..661L} obtained under Programme ID
  075.A-0603 (P.I.: Fiore),  $\mathrm{H}_2\mathrm{O}$ lines appear
coincident  in wavelength  with the  \ion{Si}{ii} $^2\mathrm{P}_{3/2}$
$\lambda1816$ line,  and affect the  \ion{Fe}{ii} $^6\mathrm{D}_{7/2}$
$\lambda$1267  and   \ion{Fe}{ii}  $^4\mathrm{F}_{7/2}$  $\lambda1659$
lines, all associated with the  $z = 3.69857$ DLA. Similarly, as shown
in  Fig. \ref{fig:grb080310},  using\texttt{ molecfit}  allowed  the location and  strength of  the $\mathrm{H}_2\mathrm{O}$
lines   affecting  the   \ion{Fe}{ii}   $\lambda2382$,  $\lambda2374$,
\ion{Fe}{ii}  $^4\mathrm{D}_{7/2}$,  and  $\lambda2389$ lines  to
be determined (amongst
others)  associated with  the $z  = 2.42743$  DLA in  the  spectrum of
\object{GRB080310} \citepads{2012A&A...545A..64D} obtained
under programme ID 080.D-0526 (PI. Vreeswijk).

\subsection{FLAMES}
%
%

\begin{figure}[ht]
  \includegraphics[width=\columnwidth]{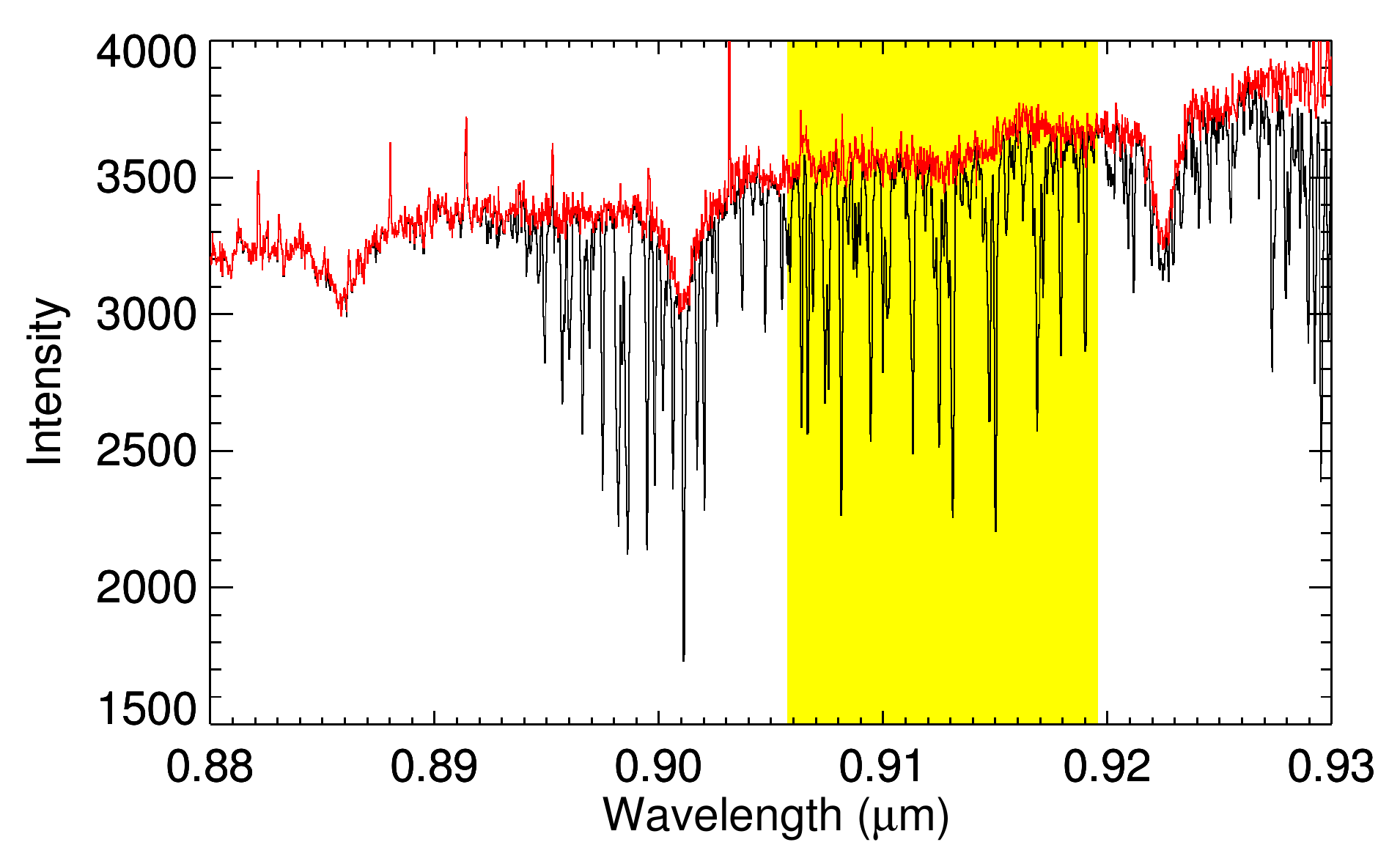}
  \caption{FLAMES/Giraffe Spectrum of  NGC~3603-117. \textit{(Black:)} Original spectrum.
    \textit{(Red:)}  telluric line  corrected spectrum.   The spectral
    range covered by the shaded  area was used to determine the amount
    of PWV. }
\label{fig:ngc3603}
\end{figure}

Figure~\ref{fig:ngc3603}  presents a Fibre  Large Array  Multi Element
Spectrograph  (FLAMES)/Giraffe spectrum of  the O4~V  star \object{NGC
  3603-117} observed with the LR8 setting and the ARGUS integral field
unit (data courtesy of M. Gieles; Prog.  ID: 079.D-0374(A)), providing a
spectral   resolving  power  of $R \approx  10\,400$.   The   presented  spectrum
corresponds to  the brightest spaxel  on the object. 

The presence of water-vapour telluric absorption lines impedes any
precise determination of the width and centroid of the Paschen
hydrogen lines at 901 and 923~nm.  However, there is usually no
observation of telluric stars for FLAMES.  \texttt{Molecfit} can be used
to determine the PWV and to correct the science spectrum by the
transmission spectrum.  As a result, a precise determination of the
width and position of the lines can be obtained by multiplying the
number of stellar lines available by three for, e.g., radial velocity
measurements  in that spectral region.

\pagebreak

\subsection{SINFONI}
\label{sec:sinfoni_j}
Archival  data   of  the  field  of  the   binary  star  \object{2MASS
  J01033563-5515561} obtained with the Spectrograph for INtegral Field
Observations       in      the      Near       Infrared      (SINFONI,
\citeads{2003SPIE.4841.1548E})  for programme ID  290.C-5022(A) (P.I.:
Delorme)  were retrieved  and reduced  with the  SINFONI  pipeline and
default  parameters.   Only  20\,s  Detector  Integration  Time  (DIT)
$J$-band  exposures obtained  at a  median airmass  of 1.28  are shown
here.   The  telluric star  \object{Hip  024337}  was  observed at  an
airmass of  1.33, approximately  66 min after  the start of  the first
science exposure.  The spectra  of the secondary star (B) and
of the telluric star were extracted using a three-pixel radius centred on
the photo-centre of each source.  Hot pixels or cosmics in the reduced
spectrum affected were manually edited out.

\begin{figure}[ht]
  \includegraphics[width=\columnwidth]{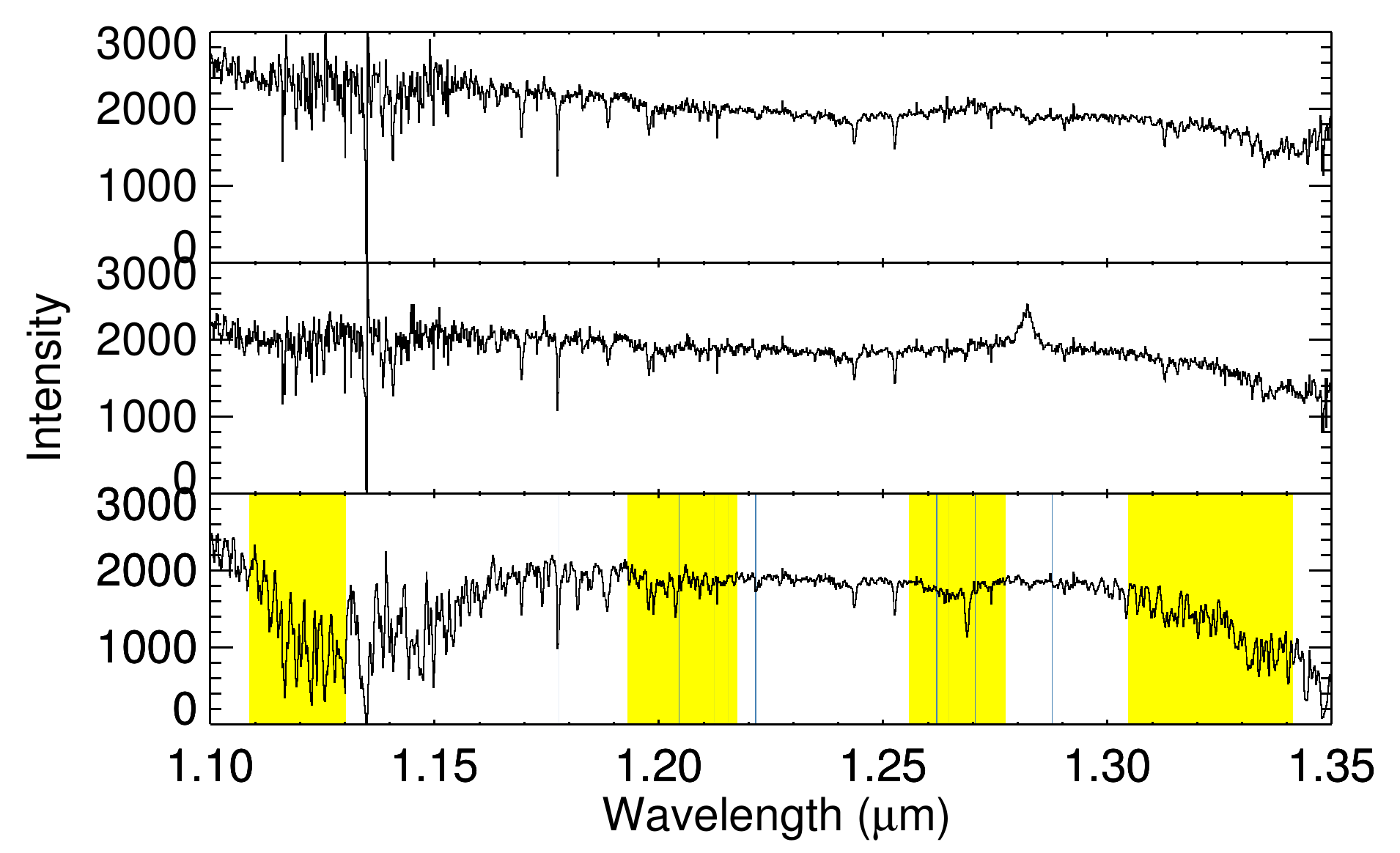}
  \caption{SINFONI $J$ band spectrum of 2MASS J01033563-5515561
    B. \textit{(Bottom:)} Pipeline reduced spectrum. The yellow shaded areas
    show the spectral ranges of the inclusion regions. Exclusion
    regions appear in blue. \textit{(Medium:)} Spectrum divided by
    the spectrum of the telluric star. \textit{(Top:)} Spectrum
    divided by the transmission spectrum calculated by \texttt{molecfit}.}
  \label{fig:sinfoni_j}
\end{figure}

Inclusion  regions were  selected to  cover telluric  lines  caused by
H$_2$O and O$_2$. Exclusion regions  were defined for the pixels affected
by cosmics. Figure~\ref{fig:sinfoni_j} shows the original spectrum and
compares  its correction  by the  transmission spectrum  calculated by
\texttt{molecfit} and by the telluric star. The spectrum corrected by the
transmission spectrum  calculated by  \texttt{molecfit} is  slightly less
noisy than  the one corrected by  the telluric star  spectrum, and it does
not show  the effect  of the \ion{H}{i}  Paschen $\beta$ line.  On the
other hand, the correction of the lines in the $\sim 1.13~\mu$m  water vapour band
is slightly less good probably because of a small change in the
line spread function.

\subsection{X-shooter}
\label{sec:xshooter}

\begin{figure}[ht]
  \includegraphics[width=\columnwidth]{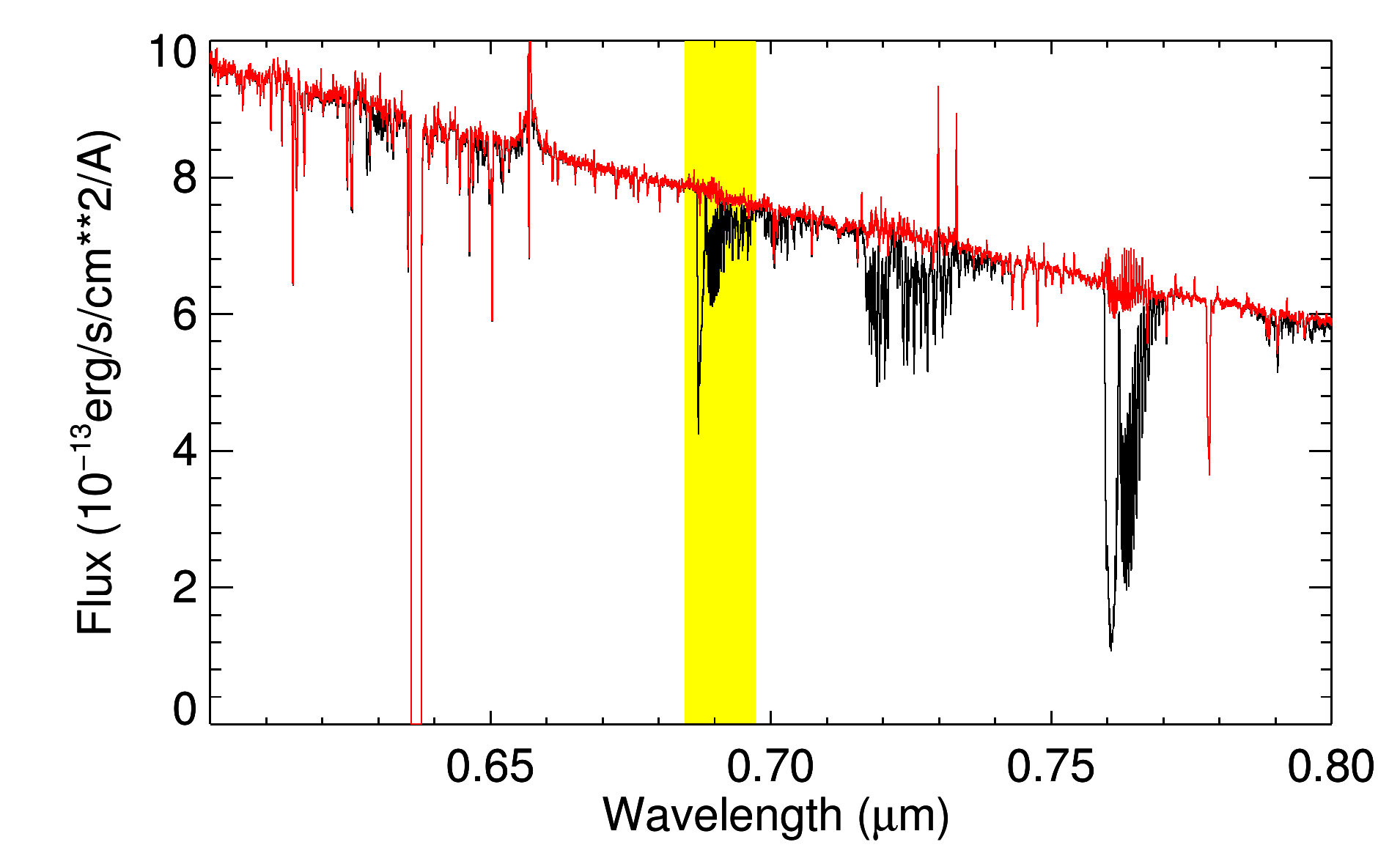}
  \includegraphics[width=\columnwidth]{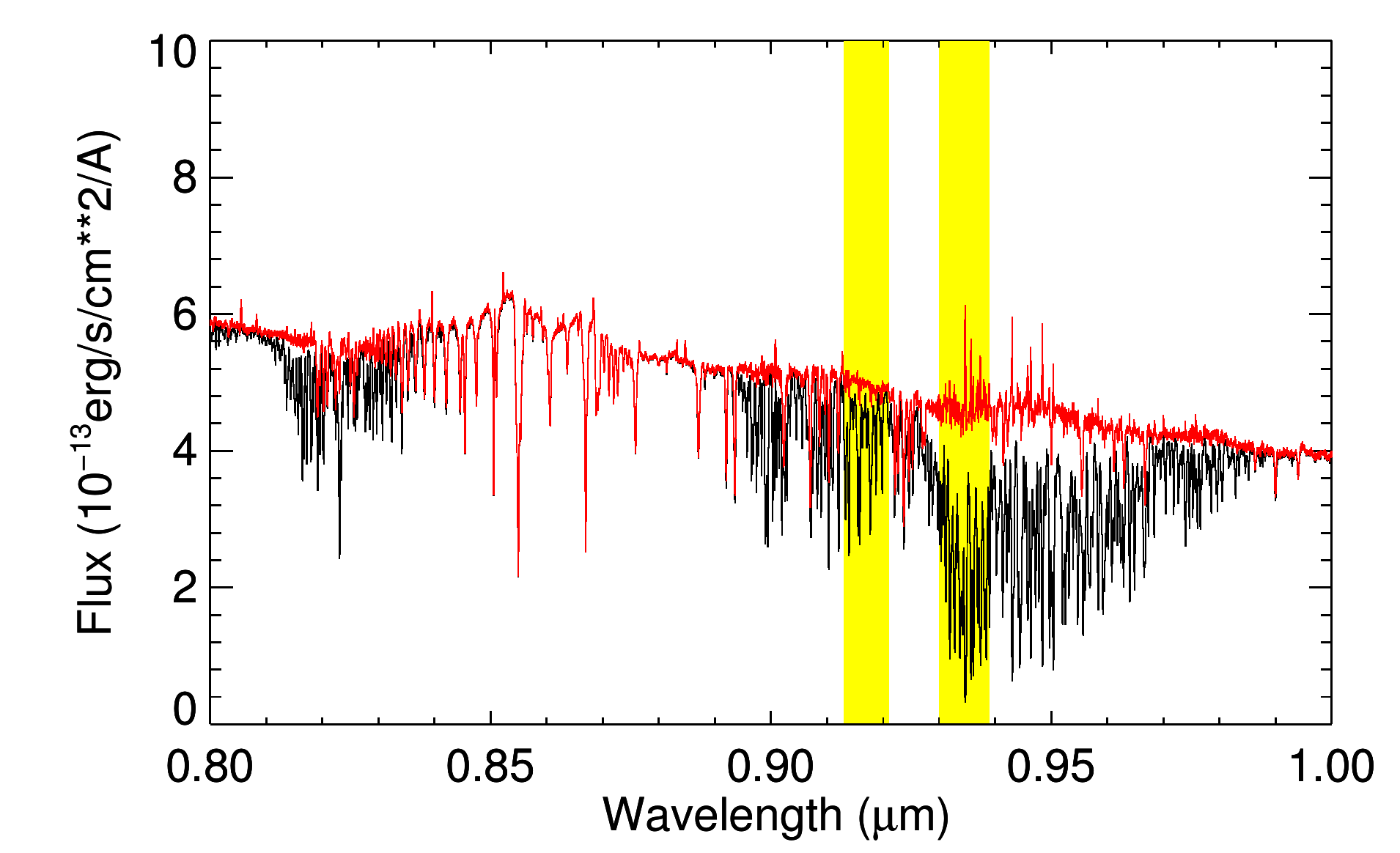}
  \caption{Visible arm X-shooter spectrum of \object{R71} LBV star.
   The pipeline reduced spectrum is shown in black. The shaded areas
   show the inclusion regions
  used to model the line spread function and determine the column
  densities of H$_2$O and O$_2$. The spectrum corrected by the derived
  transmission spectrum is shown in red.}
  \label{fig:r71}
\end{figure}

A  reduced X-shooter  \citepads{2011A&A...536A.105V}  spectrum of  the
luminous blue variable star \object{R71}  was kindly provided to us by
Andrea Mehner.  X-shooter is an  instrument composed of a  UV, a visible,
and a near-infrared  cross-dispersed  medium-resolution  spectrograph.
The data were obtained on February 1, 2014 as part of the programme ID
092.D-0024(A)  (P.I.:  Baade).    The  slit  width  used  was
$5 \arcsec \times 11 \arcsec$ and the exposure time 20 s.  Figure~\ref{fig:r71}
shows the visible part of the original spectrum together with
the  spectrum corrected  by  the transmission  spectrum calculated  by
\texttt{molecfit}, based on the fitting made in the regions marked in
in the shaded regions.

The wide slit used means that the line spread function is dominated by
the  image quality  at the  time of  the observations  (FWHM $\approx$
2\arcsec\, in this case) and not  by the slit width.  No telluric star
was observed with  the same instrumental setup because  the observation was
obtained to correct the flux loss  suffered by spectra obtained later with
a narrow slit.  Even if a telluric star  had been obtained, its
usefulness for  telluric line correction would  have strongly depended
on the stability of the seeing.

Given the overheads associated with such a short integration time, the
time spent executing  the observation of the telluric  star would have
been roughly  100\%\, of the  time spent executing the  observation of
the  science  target:  therefore,  \texttt{molecfit} also  allows  one  to
correct telluric features in spectra  obtained with a slit larger than
the image quality.

\subsection{VISIR}
\label{sec:visir}

\begin{figure}[ht]
  \includegraphics[width=\columnwidth]{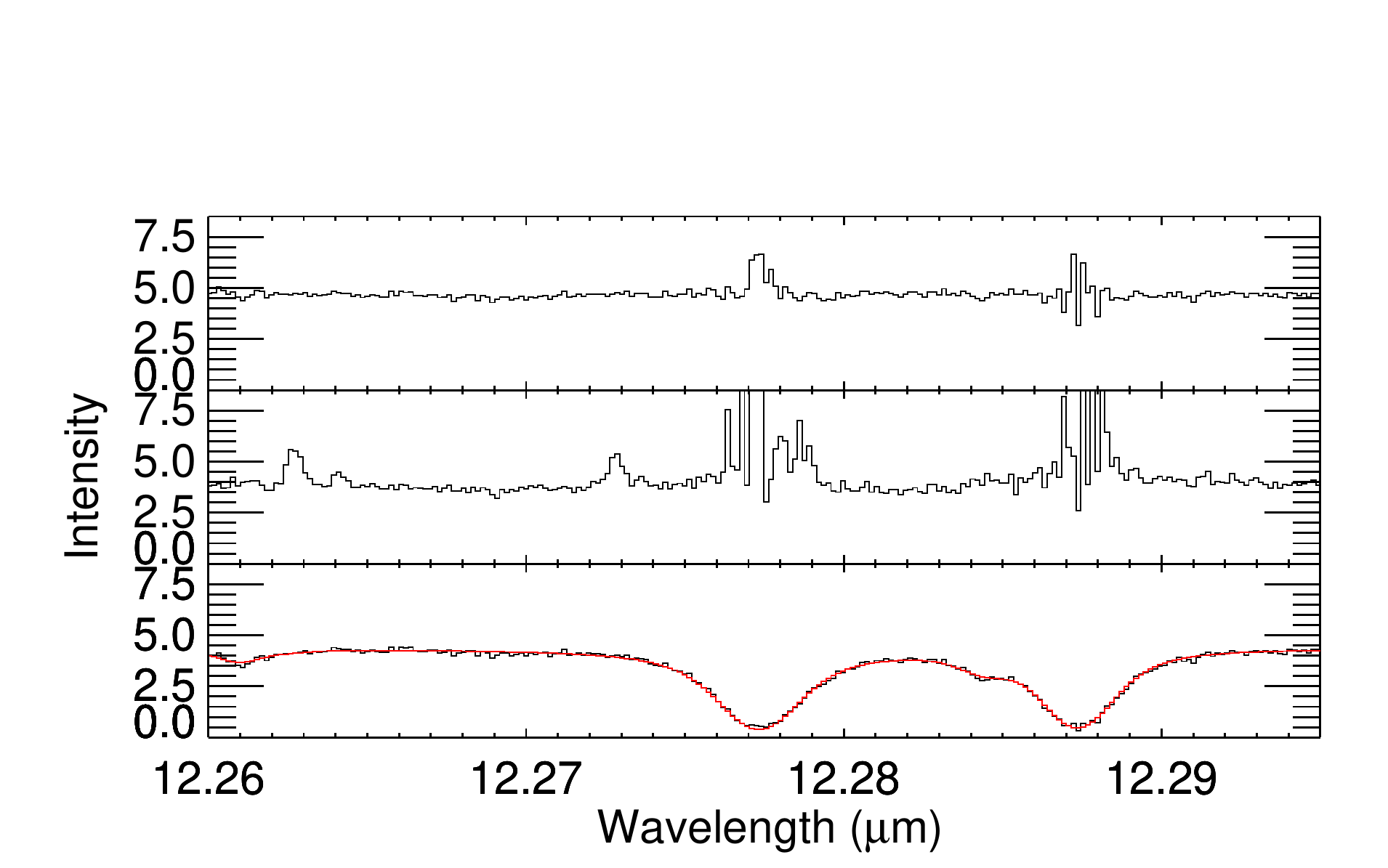}
  \caption{VISIR  high-resolution  $\lambda$12.279~$\mu$m  spectrum  of
    \object{HD 1043\,27}. \textit{Bottom:}  pipeline   reduced
    spectrum.   In  red,  the   result  of   the  fit   obtained  with
    \texttt{molecfit}. \textit{Medium:} Spectrum divided by the telluric
    star.  \textit{Top:}  Spectrum   divided  by  the  transmission
    spectrum calculated by \tt{molecfit}.}
  \label{fig:visir}
\end{figure}

Archival data of  \object{HD 104\,237} obtained with the  VLT Imager and
Spectrometer for mid InfraRed (VISIR) for programme 076.C-0129
(P.I.: van den Ancker) were  retrieved and reduced  with the VISIR
pipeline v.\ 3.5.1  and  default parameters.   Only the  12.279$\mu$m 
HRG spectrum is shown  here.  The airmass ranged from 1.68 to
1.72.

A first attempt to use \texttt{molecfit} on the spectrum of \object{HD
  104\,237}  did  not  provide  a  good correction.  Analysis  of  the
residuals  showed the  presence  of low  frequency  variations of  the
response curve, on a scale similar  to the size of the telluric lines,
and  therefore   difficult  to   correct  by  the   continuum  fitting
implemented within  \texttt{molecfit}.  Indeed, the  VISIR calibration
plan does not include any  flat fields that could correct an effect
normally taken care  of by the observation of  telluric standards.  To
alleviate this  problem, \texttt{molecfit}  was used on  the associated
telluric star  \object{HD 92\,305}, which  actually shows photospheric
absorption lines.   The same low-frequency variations can be  seen in
the spectrum of the telluric star divided by the molecfit model, which
was then fitted by a seventh-degree polynomial.  This continuum model was
then applied  to the reduced  spectrum of \object{HD  104\,237} before
another  attempt with \texttt{molecfit}.   This time  the fit  is very
satisfactory,  as can  be  seen in  Fig.\ref{fig:visir},  which can  be
compared  with  the  one for  the  same  object  shown  in Fig.  1  of
\citetads{2008A&A...477..839C}.

\texttt{Molecfit} offers the advantage of (a) less telescope time
used for observation  of a telluric star: in  this case, the execution
of the observation  for the telluric star took  36 minutes compared to
the two hours used for  the science target;  (b) the telluric  star was
observed with an airmass difference  of about 0.1, which is sufficient
to lead to significant differences  in the wings of the telluric lines
where the  molecular H$_2$ 0--0 S(2)  ($J = $4--2)  emission at 12.278
$\mu  m$ emission  line  was  expected; (c)  the  telluric star  shows
photospheric  lines that  must be  properly identified  and corrected
for.

\subsection{KMOS}
\label{sec:kmos}

\begin{figure}[t]
  \includegraphics[width=\columnwidth]{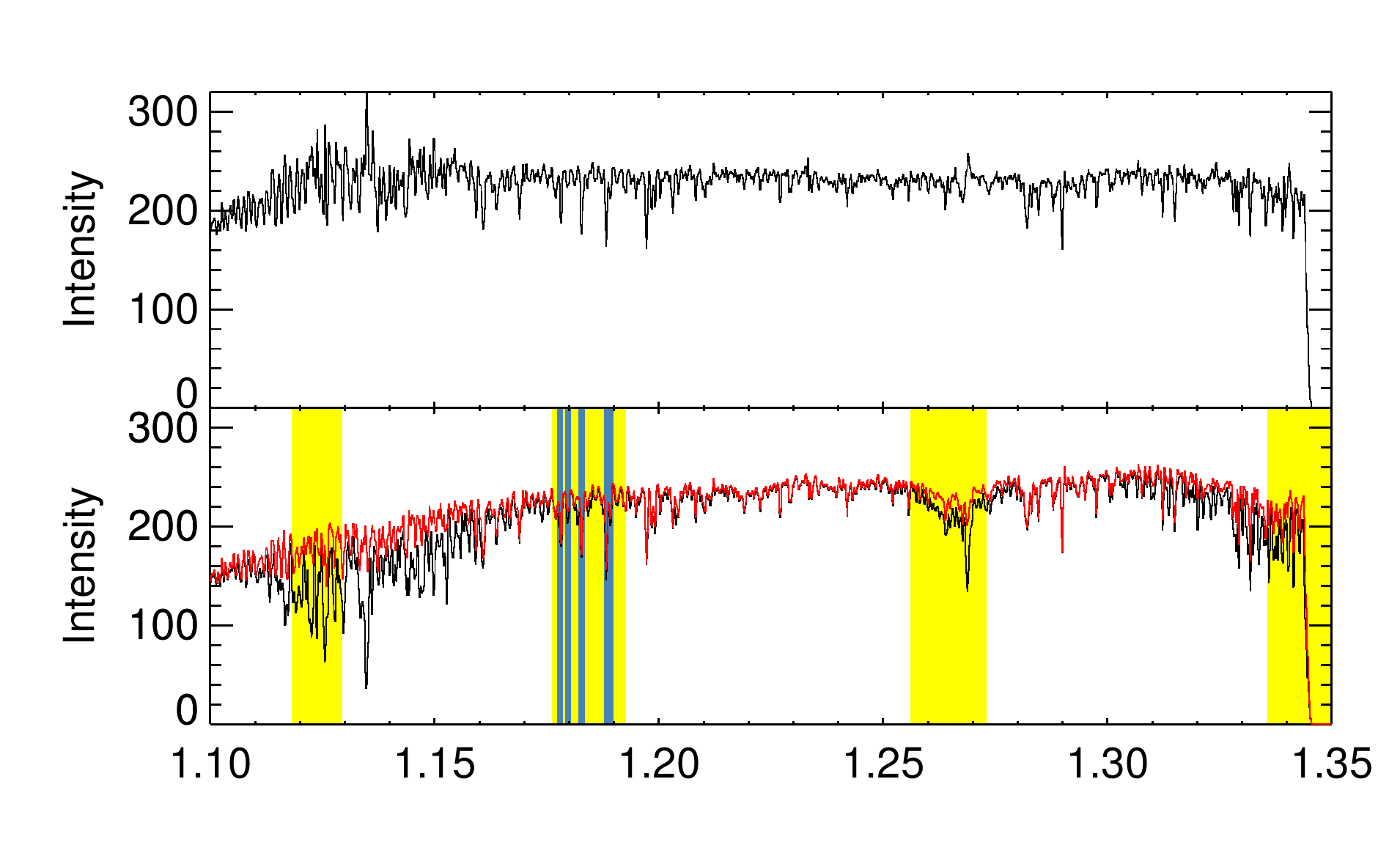}
  \includegraphics[width=\columnwidth]{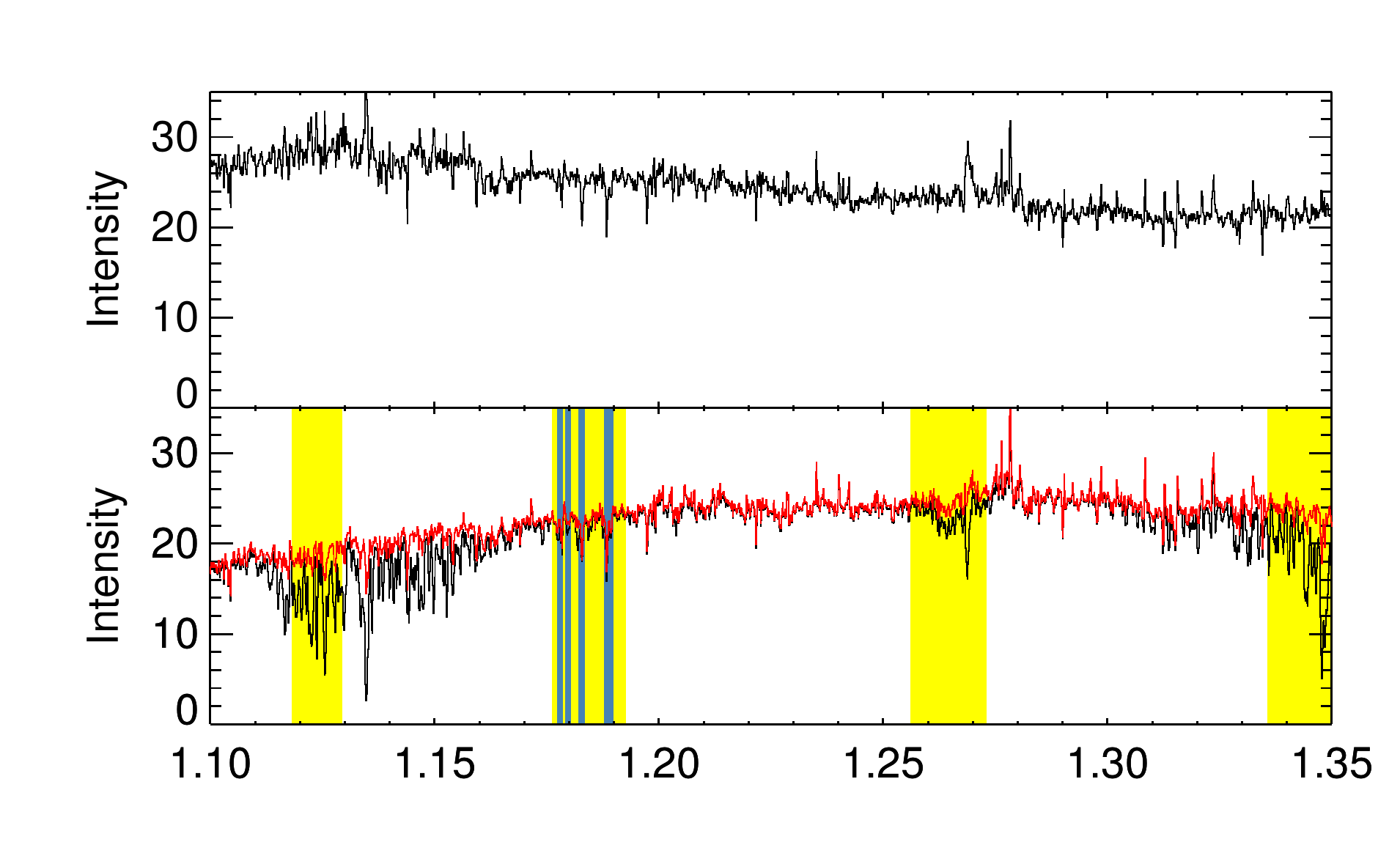}
  \includegraphics[width=\columnwidth]{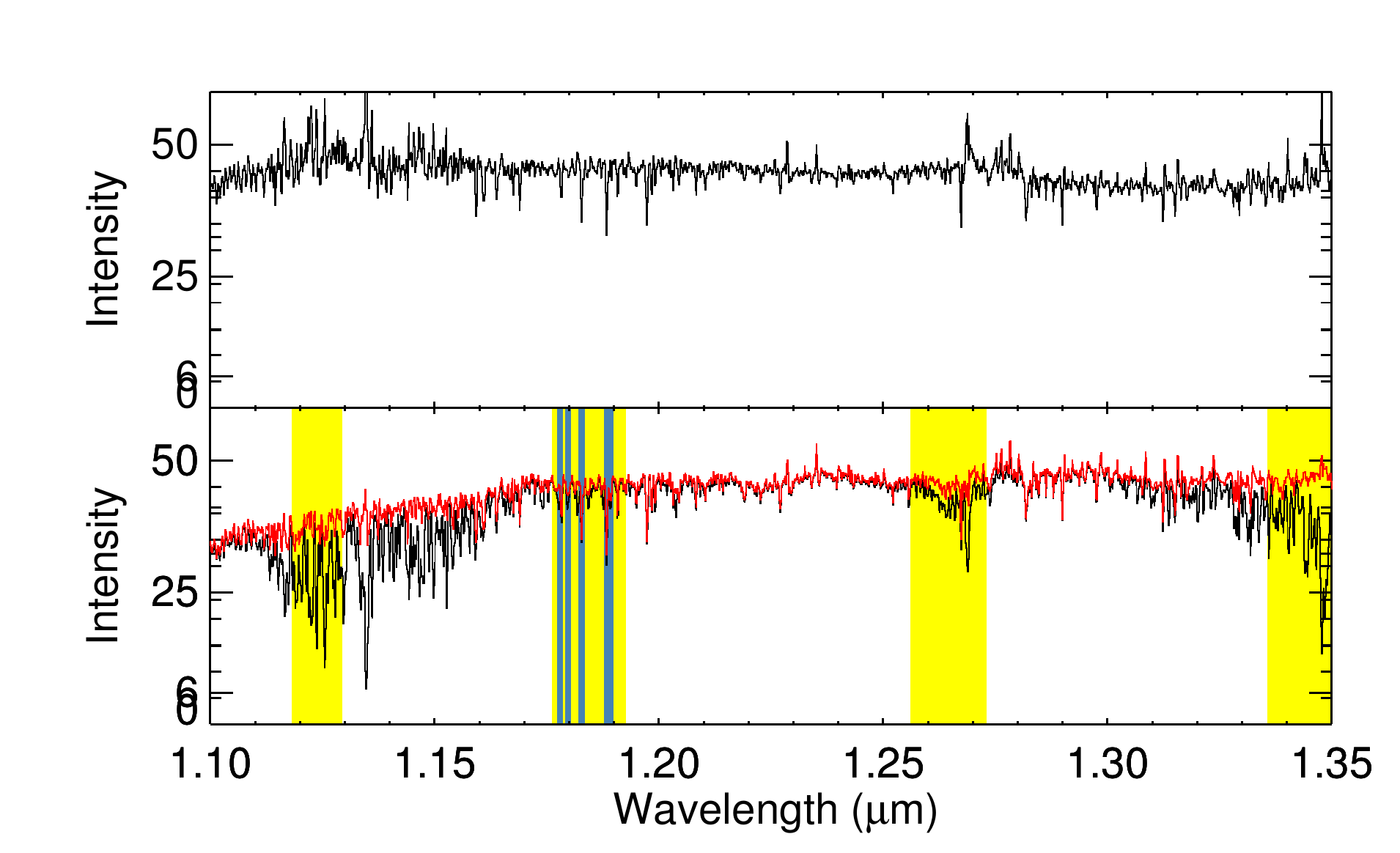}
  \caption{KMOS spectra of 3 red supergiant stars in \object{NGC~6822,}
    each feeding  a different spectrograph.  From top to  bottom: star
    \#40, star  \#36, and star \#8.  In each figure,  the bottom graph
    shows the reduced data in black, while the spectra corrected by the
    synthetic  spectrum determined  by \texttt{molecfit}  is  shown in
    red;  the top  graph shows  the  star spectrum  corrected by  the
    telluric star observed just  after.  The molecular column densities
    were determined using only the  spectrum of star \# 8: the results
    were used on the spectra of stars \# 36 and \# 40.  The line spread
    functions    were,   however,    determined    on   each    spectrum
    independently.  The  yellow shaded  areas  indicate the  inclusion
    regions, while the  blue shaded areas show the  regions excluded from
    the fit because they  correspond to spectral features intrinsic to
    the stars. }
  \label{fig:kmos}
\end{figure}

\begin{table}[h]
  \centering
  \caption{Correspondence between star ID, coordinates, KMOS arms, and spectrographs.}
  \begin{tabular}[]{ccccc}
    \hline
    \hline
    Star ID  & $\alpha(J2000)$ & $\delta(J2000)$ & Arm & Spectrograph\\
    \hline
     8       & 19 44 45.98    & $-$14 51 02.4    & 11  & 2\\
    36       & 19 44 55.93    & $-$14 47 19.6    & 1   & 1\\
    40       & 19 44 57.31    & $-$14 49 20.2    & 17  & 3\\
    \hline
  \end{tabular}
  \label{tab:kmos}
\end{table}
Reduced spectra (incl.  sky subtraction)  of three red supergiant stars of
Barnard's galaxy \object{NGC~6822}  simultaneously obtained during the
K-band         Multi         Object        Spectrograph         (KMOS,
\citetads{2013Msngr.151...21S})    science    verification   programme
60.A-9452 (P.I.: C.  Evans) were kindly provided to  us by Lee Patrick
(Patrick   et  al.,  in   preparation).  KMOS   deploys  24   arms  to
user-selected  $2.8\arcsec  \times  2.8\arcsec$  areas  at  a  spatial
sampling of  $0.2\arcsec \times 0.2\arcsec$ in  a 7.2\arcmin\ diameter
field feeding 24 integral-field units sending light to three near-infrared
medium-resolution spectrographs.   The three  stars were chosen  such that
their  spectra  were  obtained  on  different  spectrographs.  The  YJ
filter/grating  setup  was used.   See  Table  \ref{tab:kmos} for  the
correspondence    between   star    IDs,    coordinates,   arms,    and
spectrographs. For the purpose of this paper, the determination of the
column densities  for the relevant  molecules was done using  only the
spectrum of star \#8, and the determined values were then used as constant
for the two  other stars. On the other hand, the  parameters of the line
spread function were determined independently on each spectrum.

The   execution    time   required   for   the    spectra   shown   in
Fig. \ref{fig:kmos} is slightly more  than 1h. Observation of the same
telluric star  (\object{HD 187\,439} of spectral type  B6III) for each
arm was carried out right after  and took 15 min, i.e.  $\approx$ 25\%
of the time devoted to the  science. The telluric stars were chosen by
the user,  but the  difference in airmass  between the  science target
(ranging from  1.45 to 1.98) and  telluric star (ranging  from 2.00 to
2.13)  observations   make  the  corrections  by   the  telluric  star
challenging.   Instead,  \texttt{molecfit} here  again  allows one  to
avoid such problems.

Therefore  in  a  number   of  cases  \texttt{molecfit}  produce  very
satisfactory telluric corrections for KMOS. It would be interesting to
determine if the line spread function can be modelled well enough as a
function of rotation angle and temperature to avoid the need for
systematic observations of telluric stars. Such a study is outside the
scope of this paper.

\subsection{MUSE}
\label{sec:muse}

\begin{figure}[t]
  \includegraphics[width=\columnwidth]{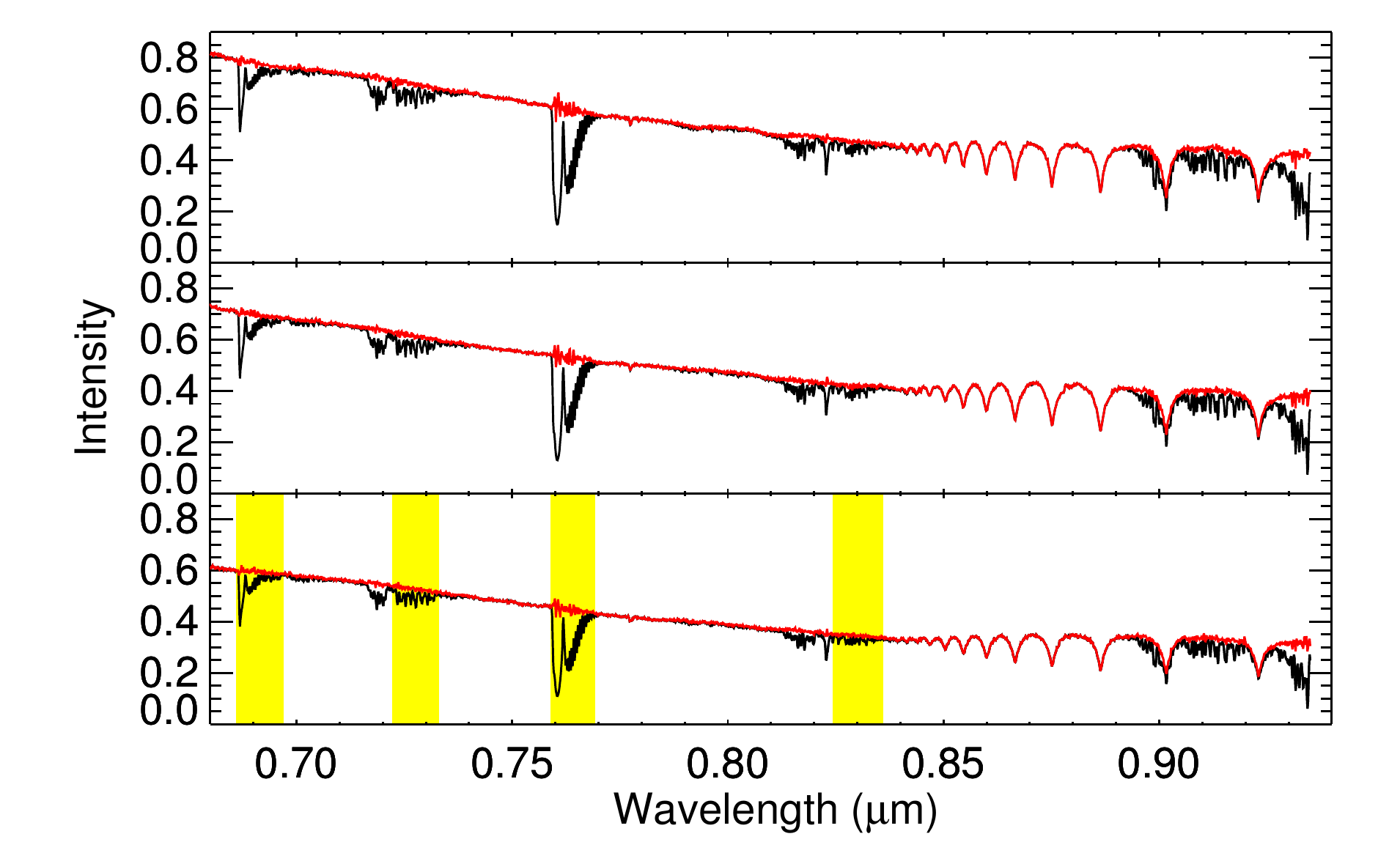}
  \caption{MUSE   spectra   of    3     stars   in the globular cluster
    \object{NGC~6\,397}.    The
    reduced data  appear in black,  while the spectra corrected  by the
    synthetic  spectrum determined  by \texttt{molecfit}  are  shown in
    red.  The molecular  column densities and line spread function were determined  using only
    star \# 5326 (\textit{bottom}) located on IFU \#7: the  results were used on the spectra
    of star \# 10327, also located on IFU \#7 (\textit{middle}), and
    \# 10893  (\textit{top}), located on IFU \# 13. The yellow shaded areas indicate the inclusion regions.
    No  regions were excluded from the fit.}
  \label{fig:muse}
\end{figure}

\begin{table}[h]
  \centering
  \caption{Correspondance between star ID, coordinates, and MUSE IFUs.}
  \begin{tabular}[]{cccc}
    \hline
    \hline
    Star ID  & $\alpha(J2000)$ & $\delta(J2000)$ & IFU \\
    \hline
    5326     & 17 40 43.51    & $-$53 40 26.0    & 7 \\
    10327    & 17 40 42.46    & $-$53 40 11.4    & 7 \\
    10893    & 17 40 40.66    & $-$53 40 13.4    & 13 \\
    \hline
  \end{tabular}
  \label{tab:muse}
\end{table}

Reduced spectra of three stars of the globular cluster \object{NGC~6\,397}
obtained   during  the   Multi-Unit   Spectroscopic  Explorer   (MUSE)
\citepads{2010SPIE.7735E..08B}   commissioning   2B   (programme   ID:
60.A-9100(C)) were kindly provided to us by Roland Bacon and Sebastian
Kamann.   MUSE  is  an  integral  field spectrograph  composed  of  24
identical   IFU  modules   sampling  a   $1\arcmin   \times  1\arcmin$
field-of-view at  ($0.2\arcsec \times 0.2\arcsec$)  spatial resolution
and up to $R \sim 3\,000$  spectral resolving power.  The three stars were
chosen such that their signal-to-noise  ratio is greater than 100 and so that
two of them fall  on the same IFU and the third  one on a different IFU.
See  Table  \ref{tab:muse} for  the  correspondance  between star  ID,
coordinate, and IFU.  The column densities for the
relevant molecules  (O$_2$, H$_2$O) and line spread  function were determined
using  only the  spectrum of  star \#5326.   The  derived transmission
spectrum was then applied to the other stars, \# 10327 and \# 10893.

The very  good wavelength calibration  and the similarity of  the line
spread  functions  between IFUs  explain  that  the same  transmission
spectrum can be used  for these IFUs.  \texttt{Molecfit} can therefore
potentially be used to correct for telluric absorption for all 90\,000
spectra obtained  by MUSE at once,  provided a suitable  star falls in
the field of view.   The spectral  resolution of the  instrument seems
low  enough not  to  be affected  by  the details  of the  atmospheric
profiles    described   in    Sect.~\ref{sec:atmprofile},    so   that
\texttt{molecfit}  can  be  used  on  a spectrum  resulting  from  the
combination  of   multiple  observations  spread   over  several,  but not
necessarily successive nights.

\subsection{CRIRES}
\label{sec:crires}

\begin{figure}[t]
  \includegraphics[width=\columnwidth]{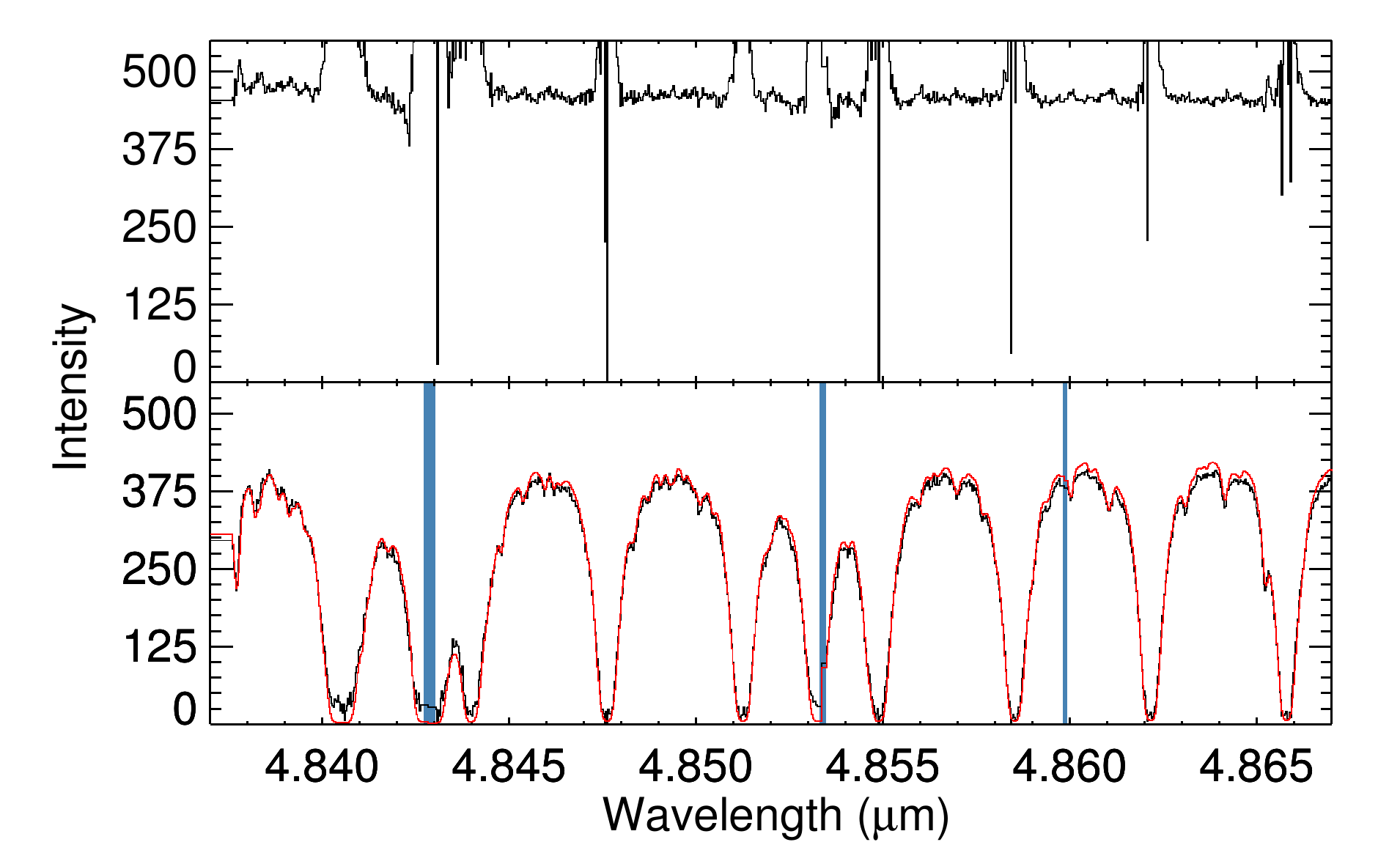}
  \includegraphics[width=\columnwidth]{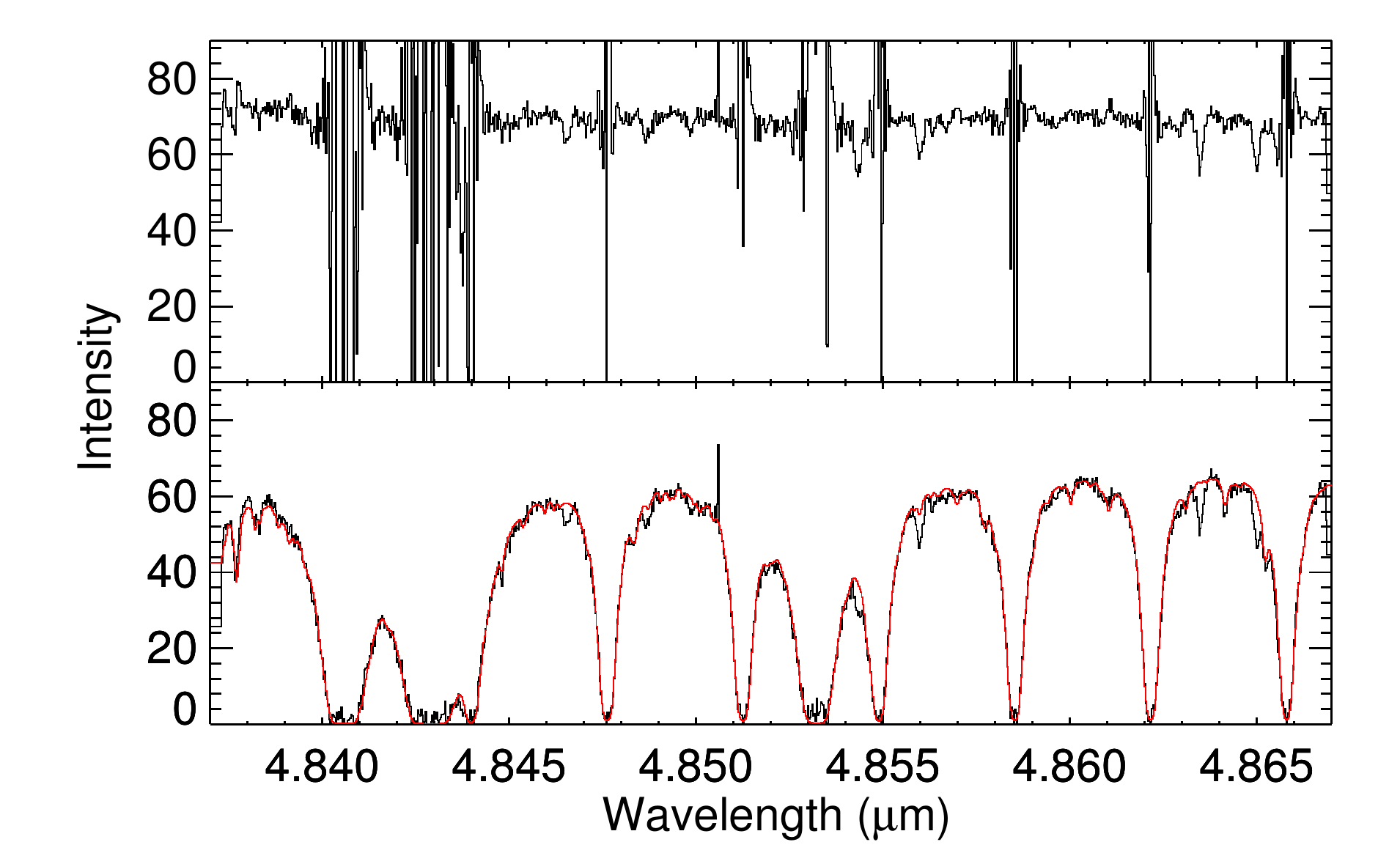}
  \includegraphics[width=\columnwidth]{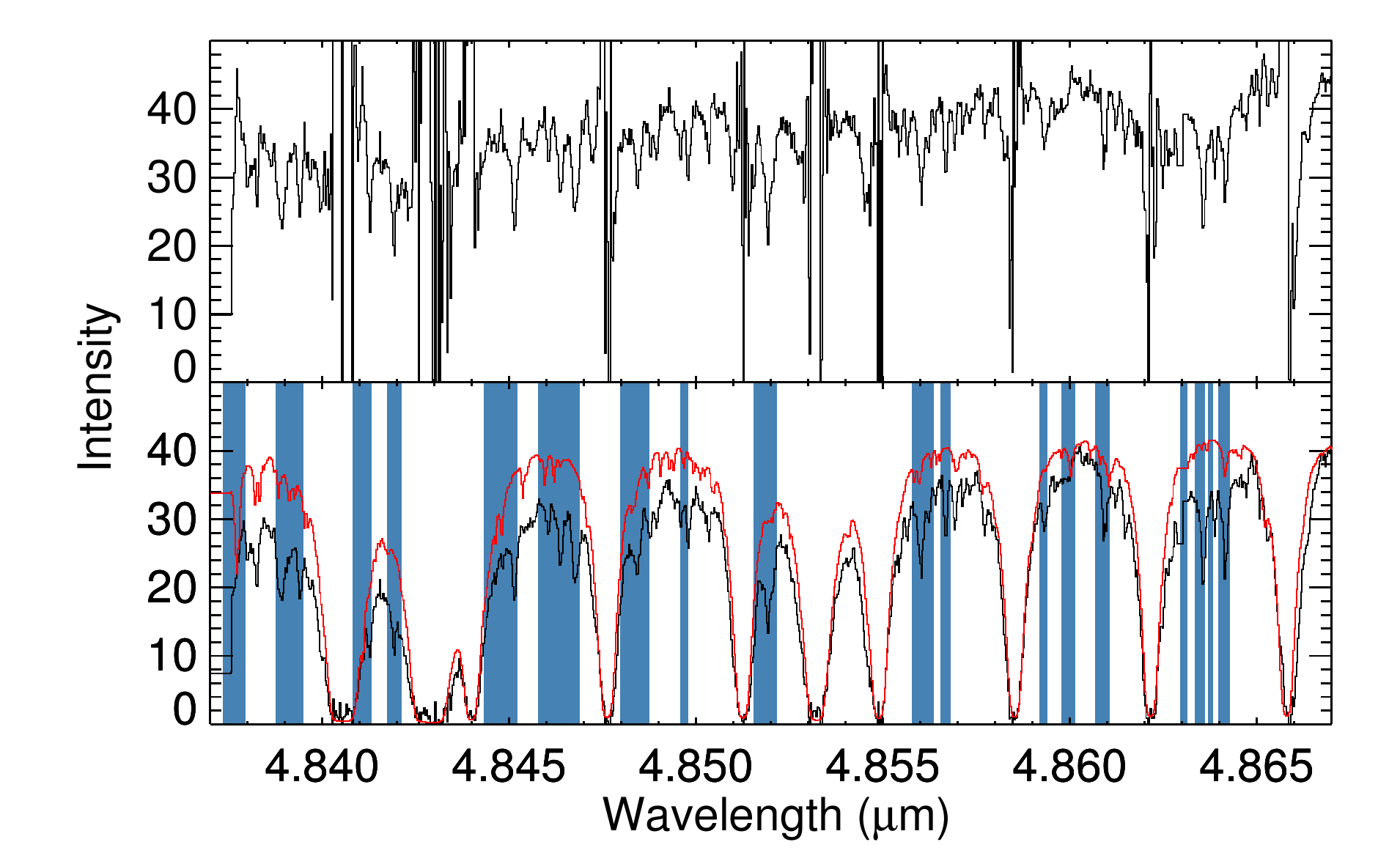}
  \caption{Extract  (detector  \#2)  of  CRIRES  $4889.5$  nm  setting
    spectra of  \object{$\gamma$ Gem} \textit{(top)}, \object{$\alpha$
      For}     \textit{(middle),}    and     \object{Barnard's    star}
    \textit{(bottom)}.  In  each  case,  the bottom  graph  shows  the
    original, reduced spectrum and the blue-shaded exclusion  regions;
    the  red spectrum  shows the  \texttt{molecfit}  derived transmission
    spectrum scaled by a constant factor. The top graph shows the {\tt
      molecfit} corrected spectrum.}
  \label{fig:crires_4889p5}
\end{figure}

Spectra  obtained  with  the  Cryogenic high-resolution  IR  \'Echelle
Spectrograph  (CRIRES, \citeads{2004SPIE.5492.1218K})  before mid-July
2014 -- when the instrument  was removed from operations to undergo an
upgrade  -- covered  a  narrow spectral  range ($\Delta\lambda  \approx
\lambda/70$) at a resolution  $R   \sim  50\,000  ~  \mathrm{to}
\,  100\,000$.   Although  the  line  spread function  is  usually  well
modelled  by a  Gaussian function  \citepads{2010A&A...524A..11S}, its
wings are more  Lorentzian. Therefore a Voigt profile  is usually best
for representing it (Villanueva  et al.  2009, internal communication), in
particular, to  attempt to  model the light  scattered on  the grating
that causes even heavily saturated lines to appear with residual light
in their  core.  The  kernel width,  on the other  hand, depends  on the
quality of the adaptive optics (AO)  correction -- when the AO is used
--  which in  turn depends  on the  `AO star'  magnitude and  on the
turbulence profile of  the atmosphere.  On the other  hand, for non-AO
observations, it may  also depend on the actual  slit width; one should
note  here  that the  slit  mechanism  lacked  reliability before  the
installation of fixed slit widths in September 2011 \citep{CRIRES_UM}.

As  mentioned  in the  CRIRES  User  Manual \citep{CRIRES_UM},  CRIRES
spectra are often affected by imprecise wavelength calibration, owing to
a lack of calibration lines.  \texttt{Molecfit} can solve this problem in
a number of  situations using the telluric lines  themselves when they
are distributed well  over the spectral  range and not  too mixed with
lines that are intrinsic to the science target.

Figure~\ref{fig:crires_4889p5} shows an extract (detector \#2 only) of
the 4889.5\,nm setting  spectra for  three stars obtained  as part  of the
CRIRES-POP  programme \citepads{2012A&A...539A.109L}:  the  A0~IV star
$\gamma$   Gem,  observed  on   October  20,   2010 (programme ID
086.D-0066, P.I.: Lebzelter);  the   F8~V  star
\object{$\alpha$  For}, observed  on  October 31,  2009 (programme ID
084.D-0912, P.I.: Lebzelter);  and the  M4~V
\object{Barnard's star},  observed on July 7, 2010 (programme ID
086.D-0066, P.I.: Lebzelter).  No telluric stars
were  observed  for this  programme.  The  number  of intrinsic  lines
increases for late-type stars, thus requiring an  increase in the
number and  size of  exclusion regions; consequently,  an F8V  star is
approximately the  latest  star for which  \texttt{molecfit}  can be
used in a non-expert mode.

\begin{figure}[h]
  \includegraphics[width=\columnwidth]{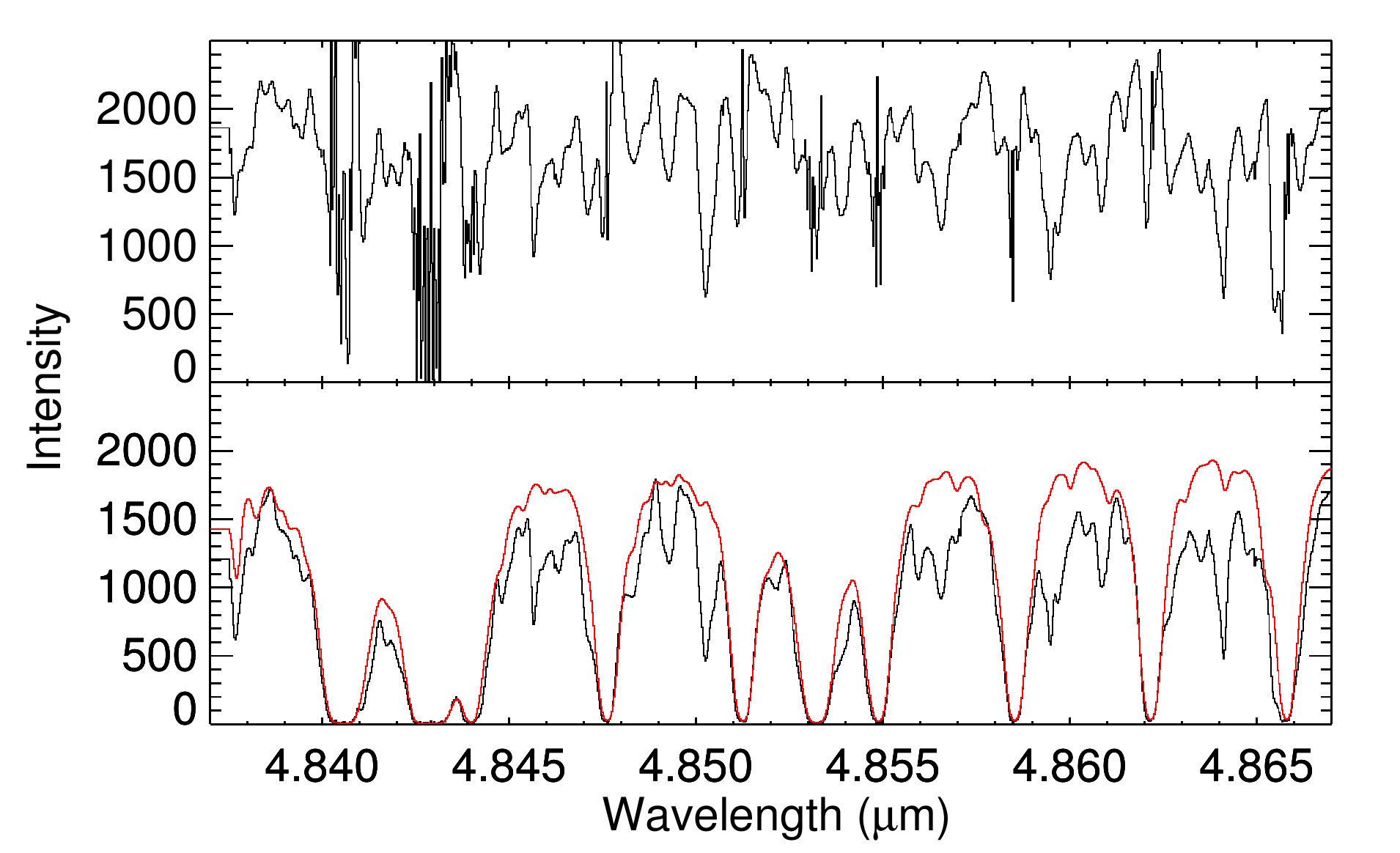}
  \caption{Extract  (detector  \#2)  of  CRIRES  $4889.5$  nm  setting
    spectra  of  \object{X  TrA};  the  red spectrum  shows  the  \texttt{molecfit}-derived transmission  spectrum scaled  by  a constant
    factor.  The   top  graph  shows  the  corrected
    spectrum. \texttt{Molecfit} was used in expert mode in this case.}
  \label{fig:crires_4889p5_e}
\end{figure}

For  later  stars, the  expert  mode (see Sect.~\ref{sec:expert})
is  required.  The  variable
carbon  star  \object{X~TrA}  was  observed  on  September  22,  2010
as part of the same programme.
Regarding the  molecular content of the atmosphere,  \texttt{molecfit} was
configured to fit only the column density for H$_2$O and CO$_2$, while
the O$_3$ and  OCS column densities were fixed.  The polynomial coefficients
for the continuum and  wavelength calibration were copied from
the results of the fit  to the \object{$\gamma$~Gem} spectrum.  The wavelength
solution found for \object{$\gamma$~Gem} was kept fixed. This method allowed one to
obtain very good results  for the telluric line correction, as shown
in Fig.~\ref{fig:crires_4889p5_e}.

\begin{figure*}[h]
  \includegraphics[]{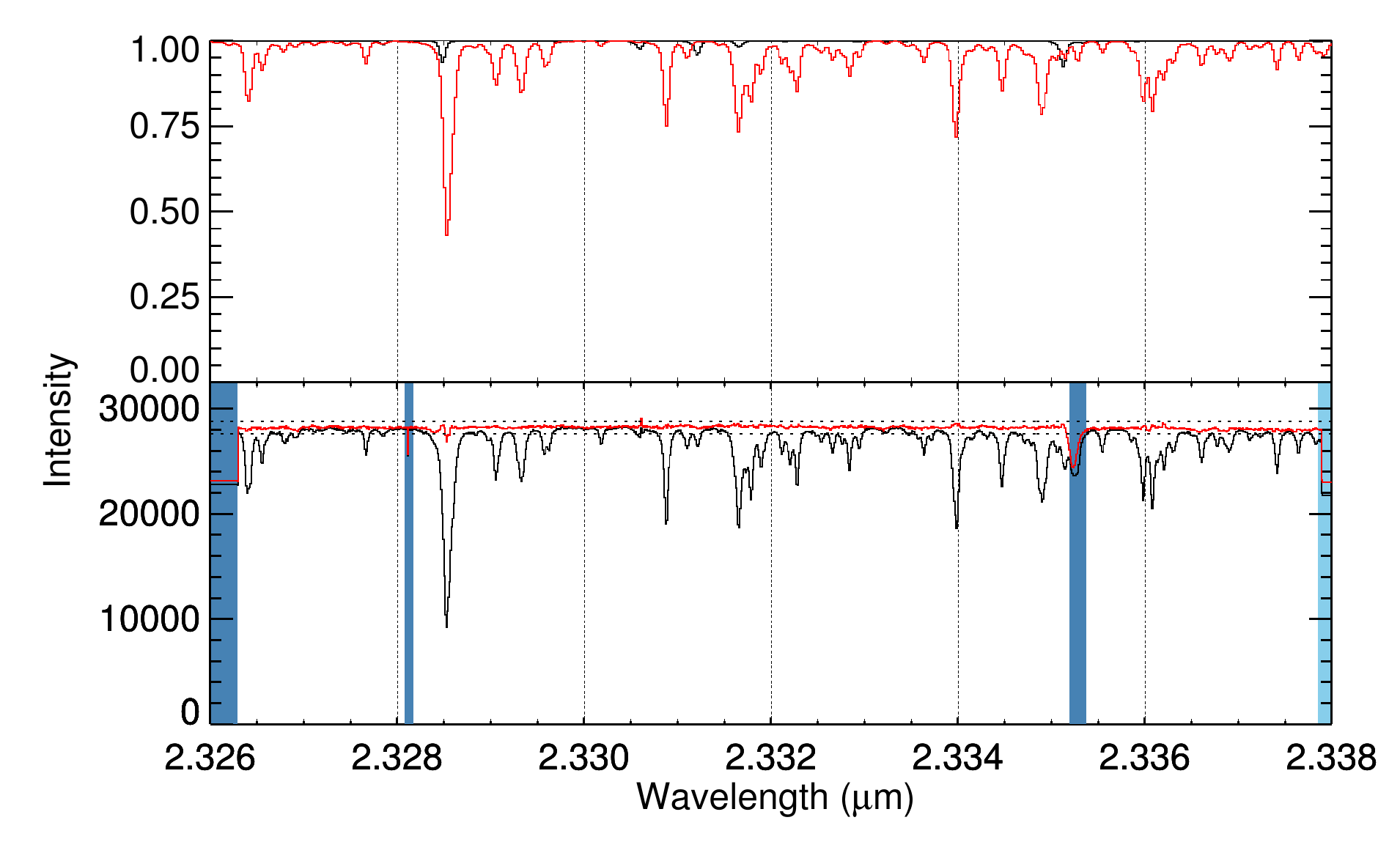}
  \includegraphics[]{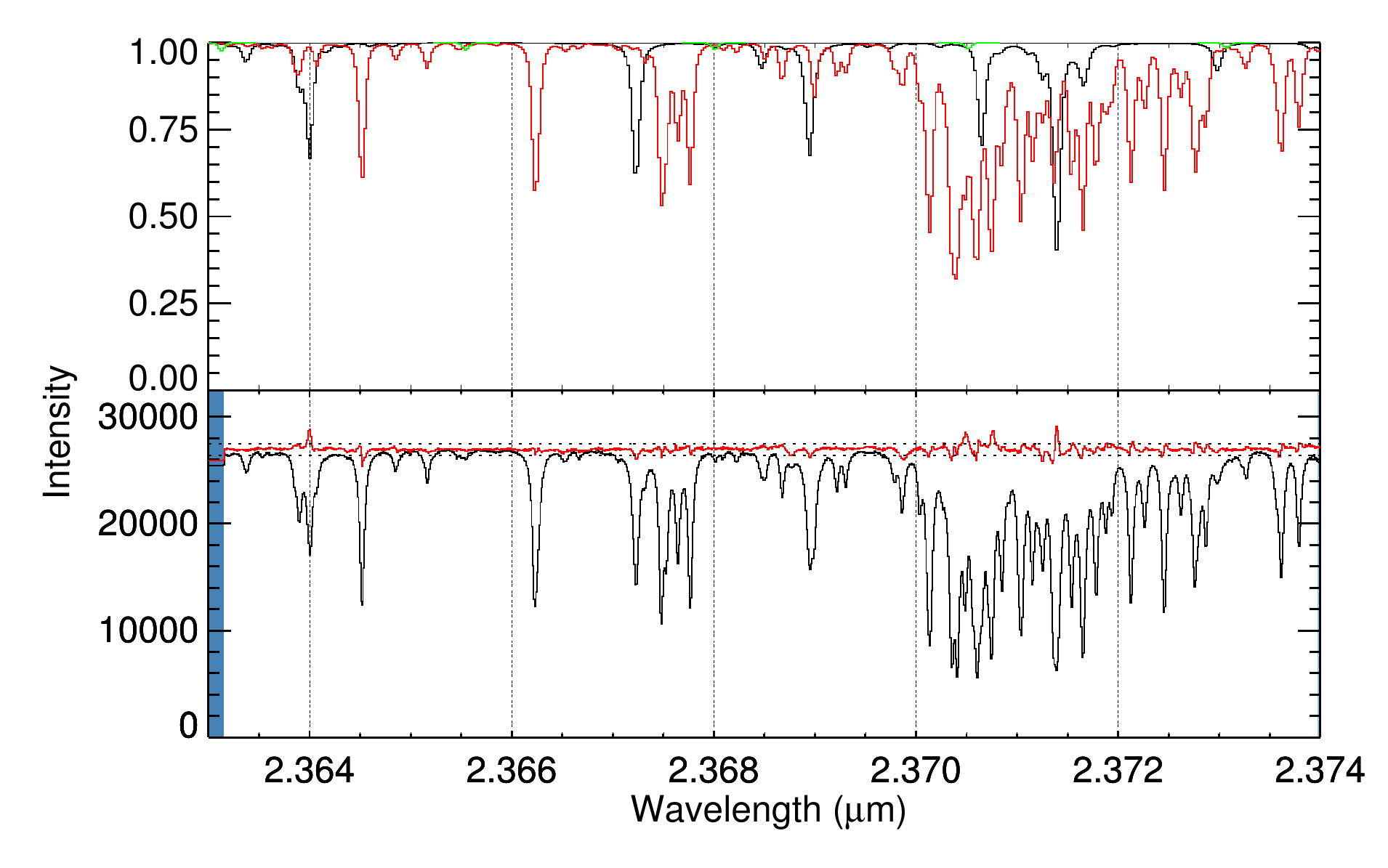}
  \caption{Extract  (detector  \#3)  of  CRIRES  $2332.2$
    \textit{(top)} 
    and $2368.7$ nm \textit{(bottom)}  setting
    spectra  of  \object{Procyon}. In each figure the  top graphs
    show generic (not model) transmission spectra for the relevant molecules:
    H$_2$O in black, CH$_4$ in red, CO$_2$ in green. 
    The  bottom   graphs  show  the original pipeline reduced spectra
    in black, as well as the molecfit-corrected
    spectra, in red. The dotted lines show 2\% deviation from the
    median value of the corrected spectra.}
  \label{fig:crires_highsnr}
\end{figure*}

To illustrate the quality of the correction on high signal-to-noise
spectra, we used \texttt{molecfit} on the 2332.2 and 2368.7 nm setting spectra
of the  F5~IV star Procyon  observed  on  November 7,   2011 (programme ID
088.D-0109, P.I.: Lebzelter). The signal-to-noise ratios - estimated 
on clean regions of the \texttt{molecfit}-corrected spectra - are
$\sim 350$ and 400, respectively. Figure~\ref{fig:crires_highsnr} 
clearly indicates a very good correction, well within 2\% except for
the core of the moderately saturated lines. A slight
difference in the quality of the correction can also be seen between 
H$_2$O and CH$_4$ lines, most likely caused by slight errors in the 
atmospheric profiles. 

Observations of telluric stars with CRIRES can easily take longer than
for the science targets. Given the high signal-to-noise of the spectra
obtained,  differences  in observing  conditions  and  in airmass  may
significantly affect spectra corrected by telluric standards.  The use
of  a   tool  such  as  \texttt{molecfit}  therefore   allows  one  to
significantly  increase  the   efficiency  of  the  instrument.   Some
programmes,  such  as  CRIRES-POP \citepads{2012A&A...539A.109L}  have
deliberately decided to only rely on synthetic telluric correction.

\subsection{Accuracy of the telluric correction}
\label{sec:summary}

To estimate  the quality  of the correction,  we compared  the standard
deviation   $\sigma$  of   the \texttt{molecfit}-corrected
spectrum  normalized by  the  continuum in  two  separate regions  for
several of the examples above: (1) a reference region with no telluric
lines and (2) a region affected by telluric lines.

The  \texttt{molecfit}-corrected   range  is  a   region  affected  by
unsaturated or moderately  saturated telluric lines  but not
covered by any inclusion region, except for VISIR and CRIRES, for which
the inclusion region corresponds  to the whole spectrum.  Both regions
are free -- as much as possible -- of  intrinsic features (either
in  emission or  absorption).  To  avoid  being  affected by
differences  in signal-to-noise ratios,  these two  ranges are  close to
each  other.   The  continuum  was  fitted  locally  by  a  low-degree
polynomial on each of the two ranges of the telluric-corrected spectrum.
Representative   values   of    $\sigma$   are   reported   in   Table
\ref{tab:accuracy}. 

We could not define reference  regions in the spectra shown in
  Fig.~\ref{fig:crires_highsnr}.    The    value   listed   in   Table
  \ref{tab:accuracy}   therefore corresponds  to  most of  the  spectra
  (avoiding an intrinsic line in  the 2332.2 setting spectrum) and can
  be compared to the signal-to-noise ratios given above.

\begin{table*}[h]
  \caption{Representative measurements of the accuracy of the telluric
    correction in spectral ranges affected by unsaturated or
    moderately saturated lines.}
  \centering
  \begin{tabular}[]{l|llc|llc}
    \hline\hline
    & \multicolumn{3}{c}{Reference} & \multicolumn{3}{c}{Telluric-corrected} \\
Instrument & $\lambda_\mathrm{min}$ &$\lambda_\mathrm{max}$& $\sigma$
&$\lambda_\mathrm{min}$ &$\lambda_\mathrm{min}$ & $\sigma$ \\
           & ($\mu m$) & ($\mu m$) & ($\times
           100)$&($\mu m$) & ($\mu m$) & ($\times 100$)\\
\hline
FLAMES              & 0.88823  & 0.89111 & 1.2 & 0.89555 & 0.89913 & 1.0\\
SINFONI             & 1.2230   & 1.2420  & 2.1 & 1.2940  & 1.3045  & 1.6\\
SINFONI             & 1.2230   & 1.2420  & 2.1 & 1.1500  & 1.1600  & 6.2\\
X-shooter VIS       & 0.878    & 0.885   & 1.0 & 0.894   & 0.890   & 1.5\\
X-shooter VIS       & 0.806    & 0.810   & 0.8 & 0.810   & 0.818   & 1.1\\
VISIR               & 12.263   & 12.273  & 2.0 & 12.283  & 12.285  & 2.5\\
KMOS \# 40          & 1.162    & 1.170   & 4.7 & 1.140   & 1.150   & 5.4\\
MUSE \# 10893       & 0.708    & 0.714   & 0.4 & 0.716   & 0.733   & 1.0\\
CRIRES $\alpha$ For & 4.8485   & 4.8500  & 2.5 & 4.8516  & 4.8524  & 2.7\\
CRIRES Procyon      &          &         &     & 2.3270  & 2.3350  & 0.6\\
CRIRES Procyon      &          &         &     & 2.3635  & 2.3740  & 1.2\\
\hline
  \end{tabular}
  \label{tab:accuracy}
\end{table*}

These  measurements show  that for  the examples  shown in  this paper,
\texttt{molecfit}  is  able  to  correct  unsaturated  lines  to  a
standard deviation better  than 2\% of the continuum.   In Paper II we
systematically investigate the quality  of the telluric corrections by
\texttt{molecfit}  over  a  large  sample of  X-shooter  near-infrared
spectra and by comparison with the IRAF \texttt{telluric} task.

\section{Conclusion}
\label{sec:conclusion}

\texttt{ Molecfit} is  a versatile tool for modelling  and correcting telluric
absorption lines.   In its most  common use, \texttt{molecfit}  fits a
spectrum  of the transmission  of the  Earth's atmosphere  over narrow,
user-selected ranges (inclusion regions)  of the spectrum of a science
target. This allows  one to  determine  the column  densities of  the
relevant molecules  causing telluric lines and the  parameters of the
line spread  function.  Alternatively, the fit can  partially or fully
make use of parameters determined on spectra of other objects obtained
at a  different time.  In  all cases, \texttt{molecfit} can  take the  best information of  temperature, pressure, and  humidity into
account in
the atmosphere above the observatory at the time of observing
the  science target.   \texttt{Molecfit} then  derives  a transmission
spectrum for  the whole range of  the input spectrum  and corrects for
it.  This  approach allows one to  regularly reach an  accuracy of the
order of 2\%  of the continuum or better  in correcting unsaturated
telluric absorption lines.

In most cases,  the input spectrum must meet  the following conditions
for  a  good  correction:  (1)  its  wavelength  calibration  must  be
accurate, (2)  the continuum  in the narrow  ranges used  as inclusion
regions is  already corrected  or can be  modelled  well by low-degree
polynomials, (3)  the inclusion regions cover unsaturated lines of
the relevant molecules whose  column density varies significantly with
time (mostly water  vapour, but also CO$_2$, O$_3$)  and whose shapes
allow  one  to determine  the  line  spread  function with  sufficient
precision.

However, the  versatility of the  tool allows several  departures from
these  conditions.   For  example,\texttt{  molecfit}  can  use  the
telluric   lines  themselves   to  provide   an   improved  wavelength
calibration,  provided that  they are  well spread  over  the spectral
range of interest and  are reasonably unaffected by intrinsic spectral
lines of  the science target.  Additionally, the user can  indicate to
\texttt{molecfit} that  absorption or emission lines  in the inclusion
regions are intrinsic to the object and exclude them from the fit.

Finally, it is important to remember that any correction method cannot
recover  with  accuracy  any  science  target  feature  coincident  in
wavelength  with  $\tau  \gtrsim  $  2 telluric  absorption  lines.   Such
\textnormal{saturated} lines  can  be easily  identified in  high-resolution
spectra  (as  obtained by  CRIRES)  where these  lines are  often
spectrally  resolved; however,  in  low-to -medium-resolution  spectra
(such  as  X-shooter),  their   shape  is  mainly  determined  by  the
instrument's spectral  resolution. In other words,  even apparently weak
telluric absorption lines can be caused by saturated lines.


\begin{acknowledgements} 

  A prototype version of \texttt{molecfit} was built around the Reference
  Forward Model, a  GENLN2-based line-by-line radiative transfer model
  originally  developed  at  the  Atmospheric, Oceanic  and  Planetary
  Physics Laboratory, Oxford University, to provide reference spectral
  calculations  for the  MIPAS launched  on the  ENVISAT  satellite in
  2002. We warmly  thank Anu Dudhia for his help  in various phases of
  the prototype  development. We also  thank Andreas Seifahrt  for his
  help in usingf LNFL/LBLRTM.

  This  study was carried  out in  the framework  of the  Austrian ESO
  In-Kind     project    funded     by    BM:wf     under    contracts
  BMWF-10.490/0009-II/10/2009  and  BMWF-10.490/0008-II/3/2011.   This
  publication is  also supported by  the Austrian Science  Fund (FWF):
  P26130  and by  the  project IS538003  (Hochschulraumstrukturmittel)
  provided by the Austrian Ministry for Research (bmwfw).

  A.G. acknowledges support from FONDECYT grant 3130361.

  We  would  like  to  thank  the  participants  to  an  internal  ESO
  mini-workshop  who  provided us  with  sample  spectra from  various
  instruments  obtained for  various scientific  purposes:  their help
  allowed us to identify a number of problems in a previous version of
  the \texttt{molecfit}  package. We are also grateful  to Andrea Mehner
  and Patrick  Lee for  providing  spectra ahead  of  publications,  to
  Annalisa De Cia for providing Fig.\,\ref{fig:grb080310}, and to Andr\'e
  M\"uller for various comments. We thank the anonymous referee whose
  constructive comments improve the quality of this article.

\end{acknowledgements}


\bibliographystyle{aa}
\bibliography{molecfit}{}


\end{document}